\documentclass[11pt,a4paper]{article}
\pdfoutput=1
\usepackage{jheppub}

\usepackage{multirow, graphicx,amssymb,url,mathrsfs,amsmath}
\usepackage{wrapfig,boxedminipage,setspace,subfigure,epsfig}
\usepackage{amsxtra,amstext,latexsym,dsfont,amsfonts}
\usepackage{color,eucal}
\usepackage[dvipsnames]{xcolor}
\usepackage{float}

\renewcommand{\d}{\delta}

\newcommand{\nn}{\nonumber}

\newcommand{\dd}{\mathrm{d}}

\newcommand{\grad}{\nabla}

\newcommand{\wh}{\widehat}

\newcommand{\tb}{\mbox{${\tilde{\beta}}$}}
\newcommand{\bb}{\mbox{${\bar{\beta}}$}}
\newcommand{\bT}{\mbox{${\bar{T}}$}}

\newcommand{\tT}{\mbox{$\tilde{T}$}}

\newcommand{\tQ}{{\tilde{Q}}}
\newcommand{\tmu}{{\tilde{\mu}}}

\newcommand{\hT}{{\wh{T}}}
\newcommand{\hm}{{\wh{\mu}}}

\title{Linear-$T$ resistivity at high temperature}

\author{Hyun-Sik Jeong,}
\author{Keun-Young Kim,}
\author{and Chao Niu}

\emailAdd{hyunsik@gist.ac.kr}
\emailAdd{fortoe@gist.ac.kr}
\emailAdd{chaoniu09@gmail.com }

\affiliation{ School of Physics and Chemistry, Gwangju Institute of Science and Technology, \\
123 Cheomdan-gwagiro, Gwangju 61005, Korea}

\abstract{\\
The linear-$T$ resistivity is one of the characteristic and universal properties of strange metals. 
There have been many progresses in understanding it from holographic perspective (gauge/gravity duality).
In most holographic models, the linear-$T$ resistivity is explained by the property of the  infrared geometry and valid at low temperature limit. On the other hand, experimentally, the linear-$T$ resistivity is observed in a large range of temperatures, up to room temperature.  By using holographic models related to the Gubser-Rocha model, we investigate how much the linear-$T$ resistivity is robust at higher temperature above the superconducting phase transition temperature. We find that strong momentum relaxation plays an important role to have a robust linear-$T$ resistivity up to high temperature. 

 }

\begin{document}

\maketitle

\section{Introduction}

One of the interesting features in strongly correlated systems is the universality in transport phenomena across very different systems. In particular, various strange metals such as cuprates, pnictides, and heavy fermions  exhibit the linear in temperature ($T$) resistivity ($\rho$)~\cite{Hartnoll:2016apf} with a remarkable degree of universality\footnote{As other examples of universal properties, there are  the Hall angle \cite{Blake:2014yla,Zhou:2015dha,Kim:2015wba,Chen:2017gsl,Blauvelt:2017koq,Kim:2010zq} at finite magnetic field and Home's law in superconductors~\cite{Homes:2004wv, Zaanen:2004aa, Erdmenger:2015qqa,Kim:2015dna,Kim:2016hzi,Kim:2016jjk}.}, 
\begin{equation}
\rho \sim T \,.
\end{equation}
It is in contrast to ordinary metals explained by the Fermi liquid theory, where $\rho \sim T^2$.

%
%
%
However, because of the difficulty in analyzing strong correlation, a complete and systematic understanding of this problem is still lacking. The gauge/gravity duality or holographic methods~\cite{Zaanen:2015oix, Ammon:2015wua,Hartnoll:2016apf} is one of the effective ways to study strongly correlated systems by mapping them  to the dual weakly interacting systems. 
%
%
%
%

In most holographic methods, the linear-$T$ resistivity is explained by the property of the  infrared (IR) geometry. See for example \cite{Charmousis:2010zz, Davison:2013txa, Gouteraux:2014hca, Kim:2015wba, Zhou:2015dha,  Ge:2016lyn, Cremonini:2016avj, Chen:2017gsl, Blauvelt:2017koq, Ahn:2017kvc}.  In these approaches one first classifies  scaling IR geometries in terms of critical
exponents such as the dynamical critical exponent ($z$), hyperscaling violating exponent ($\theta$) and
charge anomalous parameter ($\zeta$). The geometries are supported by various matter fields and couplings.
It has been shown~\cite{Donos:2014cya, Kim:2014bza,Kim:2015sma, Kim:2015wba} that the resistivity can be  computed only by horizon data, the values of metric and matter fields at the horizon $r_h$. Thus, the resistivity reads schematically $\rho \sim r_h^{f(z,\theta,\zeta)}$, where $f$ is some function of $z,\theta$, and $\zeta$. 
By considering {\it low} temperature limit, we may replace $r_h$ with $T$ by
\begin{equation} \label{rhT}
r_h \sim T^{g(z,\theta,\zeta)} \,,
\end{equation}
with a model-dependent function $g$, so 
\begin{equation} \label{rho1}
\rho \sim T^{g(z,\theta,\zeta)  f(z,\theta,\zeta)} \,. 
\end{equation}
The effect of the parameters of the system such as chemical potential ($\mu$), momentum relaxation ($\beta$), and couplings are encoded both in the critical exponents and the proportionality constants.  As a result, the resistivity is governed by the critical exponents characterizing the critical points in condensed matter systems. 

However, this approach has a limitation. The result \eqref{rho1} is  valid only at {\it small} temperature where \eqref{rhT} is justified. Mathematically, $T$ must be very small compared with any other scales in given models. For example, $T/\mu \ll 1$ and $T/\beta \ll 1$ etc.  On the other hand, phenomenologically, the linear-$T$ resistivity is observed in a large range of temperatures, up to room temperature $\sim 300 K$.  Because the phenomenological values of $\mu$ and $\beta$ are not unambiguously identified in holographic set-up, we are not sure whether the conditions $T/\mu \ll 1$ and $T/\beta \ll 1$ are sufficient to describe the linear-$T$ behavior in strange metal phase. For example, it is still possible that the strange metal regime must be realized up to $T/\mu \lesssim 1$ and the condition $T/\mu \ll 1$ is too restrictive in holographic set-up.  This question is also related with a theoretical question: how much robust is the linear-$T$ resistivity as temperature goes up? 

To investigate this issue, in this paper, i) we extend the analysis of holographic resistivity at {\it small} temperature to arbitrary finite temperature and ii) we propose a way to specify the temperature range we need to investigate.  We use an internal scale in the model, the superconducting transition (critical) temperature ($T_c$) as a reference scale.  Experimental results show that the strange metal phase with a linear-$T$ resistivity must survive up to  $T/\mu > T_c/\mu$. This condition may not be compatible with {\it small} temperature limit that most holographic methods~\cite{Davison:2013txa, Gouteraux:2014hca, Zhou:2015dha, Ge:2016lyn, Cremonini:2016avj, Chen:2017gsl, Blauvelt:2017koq, Ahn:2017kvc} rely on.

In this paper, we focus on the Gubser-Rocha model~\cite{Gubser:2009qt} and its variants~\cite{Davison:2013txa, Zhou:2015qui, Kim:2017dgz} for two reasons.
 First,  it is an interesting holographic realization of a {\it general} (non-holographic) mechanism explaining the linear-$T$ resistivity based on three conditions: i) weak momentum relaxation, which gives a connection between resistivity and shear viscosity ($\eta$), $\rho \sim \eta$, ii) the KSS (Kovtun, Son, Starinets)  shear viscosity ($\eta$) to entropy density ($s$) ratio bound i.e. $\eta \sim s$,  iii) $s \sim T$ as in the strange metal phase of the cuprates.  However, the Gubser-Rocha model has realized this mechanism only at {\it small} temperature in the sense of \eqref{rhT}.  Because of the aforementioned reason in the previous paragraph, it will be interesting to see how much robust this general mechanism is when temperature goes up. 

Second, the Gubser-Rocah model allows an analytic solution.
Note that, in the studies~\cite{Davison:2013txa, Gouteraux:2014hca, Zhou:2015dha, Ge:2016lyn, Cremonini:2016avj, Chen:2017gsl, Blauvelt:2017koq, Ahn:2017kvc}, the solutions are valid only at {\it small} temperature and cannot be used to investigate the resistivity at arbitrary temperature.  To obtain the finite temperature solutions, we should introduce certain potential terms giving asymptotically UV AdS geometry~\cite{Kiritsis:2015oxa, Ling:2016yxy, Bhattacharya:2014dea}.  In these cases, most holographic models do not allow analytic solutions at finite temperature and we should resort to numerical methods. One exception is the Gubser-Rocha model of which analytic solution has been obtained in \cite{Davison:2013txa, Zhou:2015qui, Kim:2017dgz}.

However, contrary to the previous analysis in \cite{Davison:2013txa}, where weak momentum relaxation is essential, we focus on strong momentum relaxation, which is partly inspired by \cite{Hartnoll:2014lpa}. 
In \cite{Hartnoll:2014lpa}, it is argued that if the momentum is relaxed {\it quickly}, which is an extrinsic so non-universal effect, transport can be governed by diffusion of energy and charge, which is an intrinsic and universal effect. Thus, the universality of linear-$T$ resistivity may appear in the {\it incoherent} regime (the regime of strong momentum relaxation)\footnote{There is another proposal that the linear-$T$ resistivity may appear in weak momentum relaxation regime in the case of weakly-pinned charge density waves (CDWs), where the resistivity  is governed by incoherent, diffusive processes which do not drag momentum and can be evaluated in the clean limit \cite{Delacretaz:2016ivq, Amoretti:2017axe, Amoretti:2017frz}.}.

Indeed, in this paper, we find that the linear-$T$ resistivity becomes more robust when momentum relaxation becomes stronger.  We also show that the linear-$T$ resistivity can survive  above the superconducting phase transition only when the momentum relaxation is strong enough. 
We extend our analysis further to i) higher dimensional systems in $p+1$ spacetime with $p \ge 4$ and ii)  solutions with different IR geometries~\cite{Gouteraux:2014hca}. There are two types of IR geometries depending on the strength of couplings and potentials in the action: one is conformal to AdS$_2 \times R^{p-1}$  and the other is just AdS$_2 \times R^{p-1}$. In these extended analysis, we also confirm that strong momentum relaxation enhances linear-$T$ resistivity. 

This paper is organized as follows.
In section \ref{sec2}, we  introduce the Gubser-Rocha model and its modification by the axion fields which gives momentum relaxation. In normal phase, we compute the resistivity analytically and identify the condition under which the linear-$T$ is realized. To study the temperature dependence of the resistivity above the critical temperature, we add the simplest `superconductor' sector, a massive complex scalar. 
In section \ref{sec3}, we extend the analysis in section \ref{sec2} to higher dimension and solutions with different IR geometries. 
In section \ref{sec4}, we conclude.

\section{Superconductor based on the Gubser-Rocha model} \label{sec2}

\subsection{Model: action and ansatz}

We study a 3+1 dimensional holographic superconductor model based on a Einstein-Maxwell-Dilaton-Axion theory:
\begin{equation} \label{action1}
\begin{split}
S = & \, S_1 + S_2+S_3  =  \int \dd^4x\sqrt{-g}\left( \mathcal{L}_1 + \mathcal{L}_2 + \mathcal{L}_3  \right) \,, \\ 
& \mathcal{L}_1 = R-\frac{1}{4} e^\phi F^2 -\frac{3}{2}(\partial{\phi})^2+\frac{6}{L^2}\cosh \phi  \,, \\
& \mathcal{L}_2 = -\frac{1}{2}\sum_{I=1}^{2}(\partial \psi_{I})^2 \,, \qquad
  \mathcal{L}_3 =  -|D\Phi|^2 -m^2|\Phi|^2  \,,
\end{split}
\end{equation}
where $L$ is the AdS radius, $D_{\mu} := \grad_{\mu} -iqA_{\mu}$ and the gravitational constant  is chosen to be $16\pi G = 1$.
The first term $S_1$ is the Einstein-Maxwell-Dilaton model which we call the `Gubser-Rocha model'~\cite{Gubser:2009qt}.
This model constitutes of three fields: metric, $U(1)$ gauge field, and a scalar field so called the `dilaton'.   
The metric and gauge field are minimum holographic ingredients for a quantum field theory at finite temperature and density. The dilaton was originally introduced to avoid a finite entropy at zero temperature. It turns out, with the dynamics of the dilaton, the entropy density  ($s$) becomes proportional to temperature ($T$): $s \sim T$~\cite{Gubser:2009qt}. The second term $S_2$, which is called the `axion' is added to break translation invariance so that momentum is relaxed and the resistivity is finite~\cite{Andrade:2013gsa, Zhou:2015qui,Kim:2017dgz}. In \cite{Davison:2013txa}, instead of axion, a graviton mass term \cite{Vegh:2013sk} was included which also plays a role of introducing momentum relaxation. The third term $S_3$ is the complex scalar field dual to the superconducting order parameter~\cite{Hartnoll:2008vx}\footnote{See \cite{Herzog:2014tpa} for linear-$T$ resistivity in $p$-wave holographic superconductor models without momentum relaxation.}.

The action \eqref{action1} yields the equations of motion
\begin{equation}\label{EiensteinEq}
\begin{split}
&R_{\mu\nu} -\frac{1}{2}g_{\mu\nu}\left[R-\frac{1}{4} e^\phi F^2 -\frac{3}{2}(\partial{\phi})^2+\frac{6}{L^2}\cosh \phi -\frac{1}{2}\sum_{I=1}^{2}(\partial \psi_{I})^2 -|D\Phi|^2 -m^2|\Phi|^2    \right] \\
& \qquad =\frac{1}{2}e^\phi F_{\mu\d}F_{\nu}{^\d}+\frac{3}{2}\partial_{\mu}\phi \partial_{\nu}\phi+\frac{1} {2}\sum_{I=1}^{2}(\partial_{\mu}\psi_{I}\partial_{\nu}\psi_{I})+\frac{1}{2}\left(D_{\mu}\Phi D_{\nu}^{*}\Phi^{*}+D_{\nu}\Phi D_{\mu}^{*}\Phi^{*} \right) \,, \\
%
&\grad^2\phi-\frac{1}{12}e^\phi F^2+\frac{2}{L^2} \sinh (\phi) =0 \,, \\
&\grad_{\mu}(e^\phi F^{\mu\nu})-iq\Phi^{*}(\partial^{\nu}-iqA^{\nu})\Phi+iq\Phi(\partial^{\nu}+iqA^{\nu})\Phi^{*}=0  \,, \\
&\grad^{2}\psi_{I}=0 \,, \qquad
D^{2}\Phi-m^2\Phi=0  \,,
\end{split}
\end{equation}
and we use the following ansatz:
\begin{equation} \label{ansatz1}
\begin{split}
&\dd s^2 =  g_{tt} \dd t^2  + g_{zz} \dd z^2 + g_{xx} \dd x^2 + g_{yy}\dd y^2 \\
& \quad \ \, = \frac{L^2}{\tilde{z}^2} \left[-(1-\tilde{z})U(\tilde{z}) \dd \tilde{t}^2+\frac{\dd \tilde{z}^2}{(1-\tilde{z})U(\tilde{z})}+V(\tilde{z})\dd \tilde{x}^2 +V(\tilde{z}) \dd \tilde{y}^2 \right] \,, \\
&A=L(1-\tilde{z})a(\tilde{z}) \dd \tilde{t} \,, \quad  \phi=\frac{1}{2} \log[1+\tilde{z}\,\varphi(\tilde{z})] \,, \quad \Phi=\tilde{z}\, \eta(\tilde{z}), \\
&\psi_{1}=\tilde{\beta} \, \tilde{x} \,, \quad \psi_{2}=\tilde{\beta} \, \tilde{y} \,,
\end{split}
\end{equation}
where 
\begin{equation} \label{tildes}
\tilde{z} :=  \frac{z}{z_{h}} \,, \qquad \tilde{t} :=\frac{t}{L^{2} z_{h}} \,, \qquad \tilde{x} :=\frac{x}{L^{2} z_{h}} \,, \qquad \tilde{y} :=\frac{y}{L^{2} z_{h}} \,, \qquad  \tb :=  \beta \, z_{h} \,.  
\end{equation}
Here, we choose the holographic coordinate $\tilde{z}$ such that the black hole horizon is located at $\tilde{z}=1$ and the boundary is at $\tilde{z}=0$.  
Our coordinate system is related to \cite{Zhou:2015qui,Kim:2017dgz} by $z = 1/r$ and the specific form of ansatz \eqref{ansatz1} is chosen for convenience in numerical analysis for superconducting phase.  For simplicity, we set $L=1$  from here on.

Suppose that the IR physics of a system is well described by hydrodynamics with a minimal shear viscosity ($\eta \sim s$), which is typical in  strongly correlated systems with holographic duals.  If this system lose momentum {\it weakly} by coupling to random disorder the resistivity turns out to be proportional to viscosity. i.e. $\rho \sim \eta \sim s$. Finally, for a system with $s \sim T$ such as the strange metal phase  of the cuprates, $\rho \sim T$ \cite{Davison:2013txa}. The Gubser-Rocha model is a holographic realization of this mechanism although  it has a dynamical exponent $z \rightarrow \infty$ and a hyperscaling violating exponent $\theta \rightarrow -\infty$ with the fixed $\theta/z = -1$.

\subsection{Linear-$T$ resistivity in strange metal phase}
For normal phase, $\Phi=0$ ($\eta(\tilde{z})=0$), the analytic solution is available~\cite{Zhou:2015qui,Kim:2017dgz}
\begin{equation} \label{sol01}
\begin{split}
U(\tilde{z})&=\frac{1+(1+3\tQ)\tilde{z}+\tilde{z}^2(1+3\tQ(1+\tQ)-\frac{1}{2}\tb^{2})}{(1+\tQ \tilde{z})^{3/2}} \,, \qquad
V(\tilde{z})=(1+\tQ \tilde{z})^{3/2} \,, \\
a(\tilde{z})&=\frac{\sqrt{3\tQ(1+\tQ)\left(1-\frac{\tb^2}{2(1+\tQ)^2} \right)}}{1+\tQ \tilde{z}} \,, \qquad \varphi(\tilde{z})=\tQ \,,
\end{split}
\end{equation}
where $\tQ$ is a parameter which will be expressed in terms of physical parameters such as temperature ($T$), chemical potential ($\mu$) or momentum relaxation parameter ($\beta$). 
The temperature and chemical potential read 
\begin{align}
&T= \frac{g_{tt}'(z)}{4\pi \sqrt{g_{tt} g_{zz}}}\Bigr|_{z_{h}} = \frac{1}{z_h}\frac{6(1+\tQ)^{2}-\tb^{2}}{8\pi(1+\tQ)^{3/2}} =: \frac{1}{z_h} \tT \,,  \label{tTeq} \\
&\mu= A_t(0) = \frac{1}{z_h} \sqrt{3\tQ(1+\tQ)\left(1-\frac{\tb^{2}}{2(1+\tQ)^{2}}\right)}  =: \frac{1}{z_h} \tmu  \label{tmueq}\,,
\end{align}
where the last equalities define $\tT = T z_h$ and $\tmu = \mu z_h$. The conductivity can be computed~\cite{Donos:2014cya, Blake:2013bqa, Kim:2015wba, Kim:2015sma} as
\begin{equation} \label{conduct1}
\sigma_{DC} := e^{\phi} - \frac{A'^{2} \, g_{xx} \, e^{2\phi}}{\beta^2 \, g_{tt} \, g_{zz}}\Bigr|_{z\rightarrow z_{h}} =  \sqrt{1+\tQ} + \frac{\sqrt{1+\tQ}}{(\tb/\tmu)^2} \,,
\end{equation}
%
where $\tQ$ may be a function of  $\tT, \tb$ or $\tmu, \tb$ from \eqref{tTeq} or \eqref{tmueq}.  

\begin{figure}[]
 \centering
     {\includegraphics[width=7.3cm]{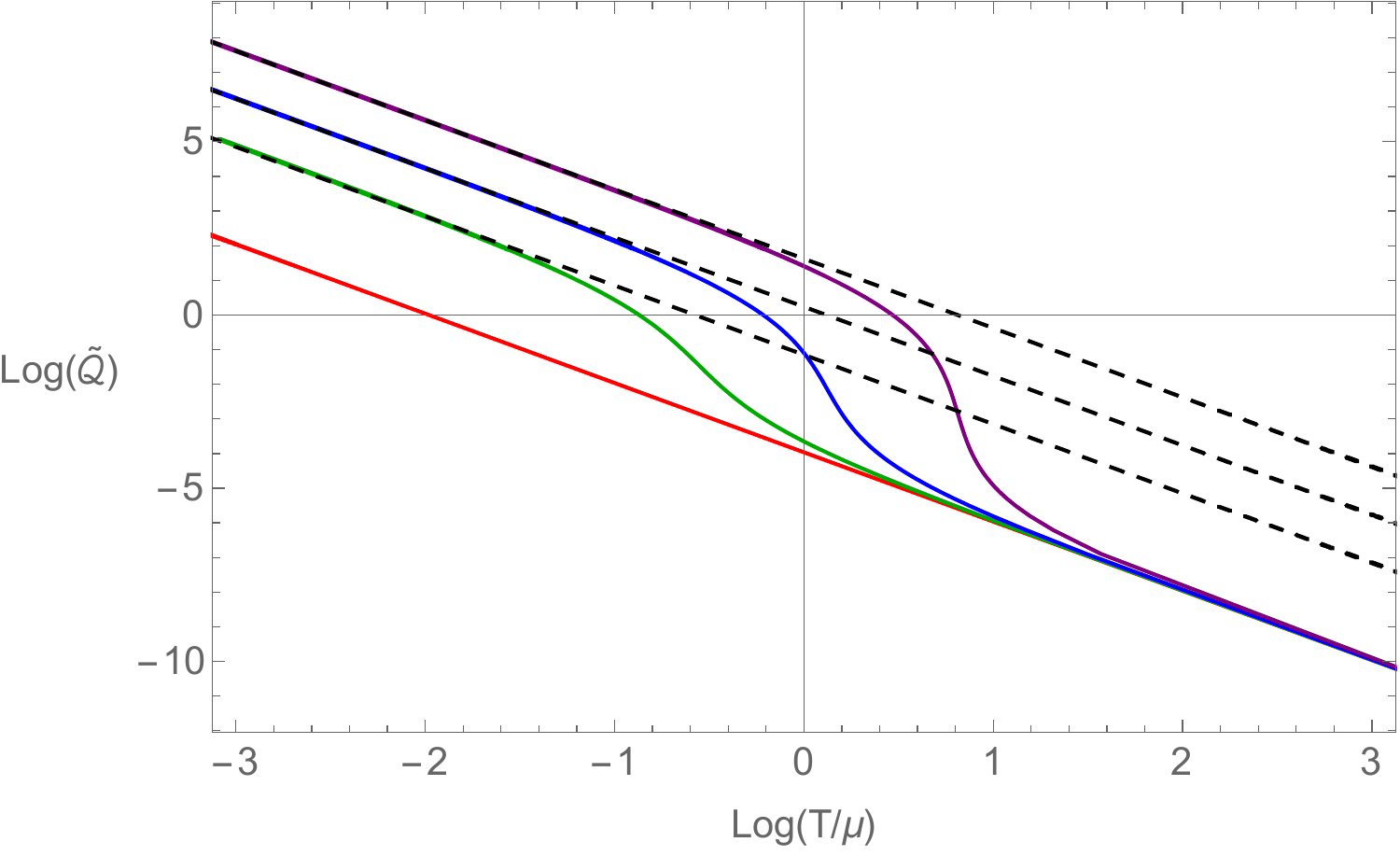} \label{}} 
          \caption{$\log \tQ$ vs $\log \bT$. The solid curves correspond to $\bb= 0.1, 5, 10, 20$ (Red, Green, Blue, Purple). The black dashed line represents \eqref{tQ3}.}\label{fig4}
\end{figure}

Variables with tilde are scaled by $z_h$ and convenient for numerical analysis. We want to fix chemical potential so express the conductivity in terms of $T$ and $\beta$ at fixed $\mu$. For this purpose we define 
\begin{align}
&\bT := \frac{T}{\mu} = \frac{\tT}{\tmu} =  \frac{6(1+\tQ)^{2}-\tilde{\beta}^{2}}{4\sqrt{6}\pi\sqrt{\tQ(1+\tQ)^{2}(2(1+\tQ)^{2}-\tilde{\beta}^{2})}} \,,  \label{bteq} \\ 
&\bb := \frac{\beta}{\mu} = \frac{\tb}{\tmu} =  \sqrt{\frac{2(1+\tQ)\tilde{\beta}^{2}}{3\tQ(2(1+\tQ)^2-\tilde{\beta}^2)}} \,, \label{bbeq}
\end{align}
where we used \eqref{tildes}, \eqref{tTeq} and \eqref{tmueq}. By combining \eqref{bteq} and \eqref{bbeq} we can obtain $\tQ$ as a function of $\bT$ and $\bb$ analytically, i.e. $\tQ(\bT, \bb)$. Thus, the electric conductivity \eqref{conduct1} can be expressed in terms of $\bT$ and $\bb$. However, because the analytic expression of $\tQ(\bT, \bb)$ is too complicated and not so illuminating, we do not show it here. Instead, we display its plots in Fig. \ref{fig4} and report its asymptotic form, \eqref{tQ1} - \eqref{tQ3}, at some limits which are relevant for our study.  

Because we are interested in the temperature dependence of $\tQ$ we make a log-log plot ($\log \tQ$ - $\log \bT$) at fixed $\bb$ to read off the power of $\bT$ in Fig. \ref{fig4}.
The solid curves correspond to $\bb= 0.1, 5, 10, 20$ (Red, Green, Blue, Purple).  Interestingly, the power turns out to be the same, $\tQ \sim \bT^{-2}$, at both small $\bT$ and large $\bT$ while there is a change in the middle. This change is bigger for large $\bb$. Fig. \ref{fig4} can be understood by the following analytic expansions of $\tQ(\bT, \bb)$.
\begin{align}
&\tQ \sim \frac{3(1+\bb^2)^{2}}{8\pi^{2}(2+3\bb^2)\bT^2}  \,, \qquad (\bT \ll 1) \,, \label{tQ1}\\ 
&\tQ \sim \frac{3}{16\pi^{2}\bT^2}  \,, \, \ \quad  \qquad \qquad (\bT \gg 1 \ \ \mathrm{or} \  \ \bb \ll 1) \,, \label{tQ2} \\
&\tQ \sim \frac{\bb^2}{8\pi^2 \bT^2} \,, \qquad \qquad \qquad (\bb \gg 1) \,. \label{tQ3}
\end{align}
Eqs. \eqref{tQ1} and \eqref{tQ2} agree with Fig. \ref{fig4} at small $\bT$ and large $\bT$. 
In $\bT \ll 1$ limit, the slopes are the same, but the starting points are varying with $\bb$, as shown in \eqref{tQ1}. In $\bT \gg 1$ limit, all curves coincide regardless of $\bb$ as shown in \eqref{tQ2}. 
Another interesting limit is the strong momentum relaxation limit $\bb \gg 1$, \eqref{tQ3}, which is displayed as the dashed lines in Fig. \ref{fig4}. Note that Eq. \eqref{tQ3} is valid for a larger range of $\bT$ for a bigger $\bb$, which will play an important role in our later discussion for linear-$T$ resistivity.

In the above limits, \eqref{tQ1}-\eqref{tQ3}, the conductivity \eqref{conduct1} behaves as
\begin{align}
&\sigma_{DC} \sim \sqrt{\tQ}\left(1+\frac{1}{\bb^2}\right)  \sim \frac{\sqrt{3}\left(1+\bb^2\right)^2}{2 \pi \bb^2 \sqrt{4+6 \bb^2}}\frac{1}{\bT}  \,, \qquad   (\bT \ll 1) \,, \label{sigmasmallT} \\
&\sigma_{DC} \sim 1+\frac{1}{\bb^2}  \,, \qquad \qquad \qquad \qquad \qquad \qquad \qquad \ \  (\bT \gg 1)  \label{sigmabigT}
\end{align}
 for given $\bb$ and 
\begin{align}
& \sigma_{DC} \sim \frac{1}{\bb^2}\sqrt{1+\tQ} \sim  \frac{1}{\bb^2}\sqrt{\left(1+\frac{3}{16 \pi^2 \bT^2}\right)} \,, \qquad (\bb \ll 1) \,, \label{sigmasmallb} \\
&\sigma_{DC} \sim  \sqrt{\tQ} \sim \frac{\bb}{2\sqrt{2}\pi \bT}  \,, \qquad \qquad \qquad  \qquad \qquad  \ \ (\bb \gg 1) \label{sigmabigb} 
\end{align}
for given $\bT$. 
We find that the resistivity ($\rho = 1/\sigma_{DC}$) is linear to temperature in two cases: $\bT \ll 1$ \eqref{sigmasmallT} and $\bb \gg 1$ \eqref{sigmabigb}. The former has something to do with a result in \cite{Davison:2013txa} and the latter is one of our main results. 

\begin{figure}[]
 \centering
     \subfigure[$\bb=0.1$]
     {\includegraphics[width=7.3cm]{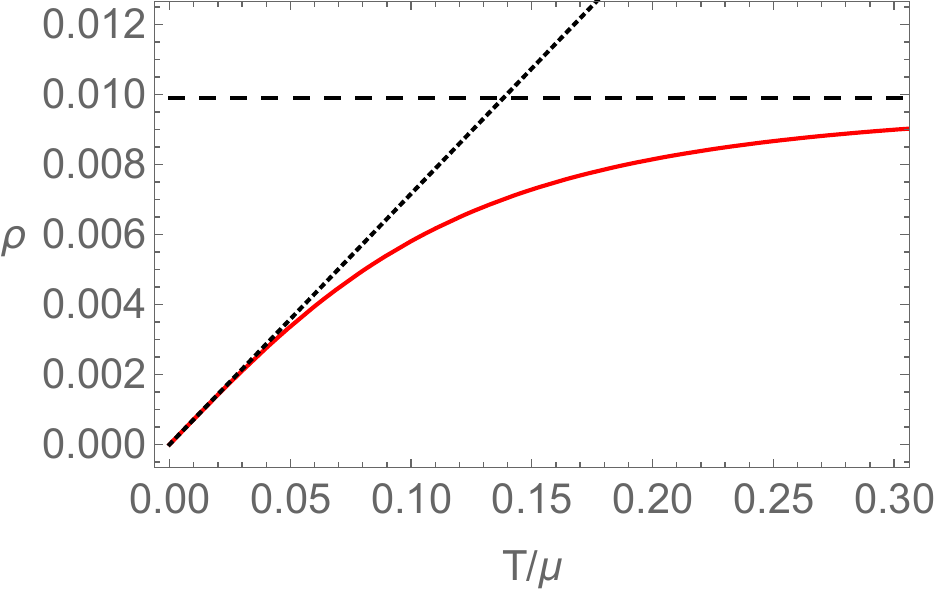} \label{}}
 \subfigure[$\bb=1$]
     {\includegraphics[width=7cm]{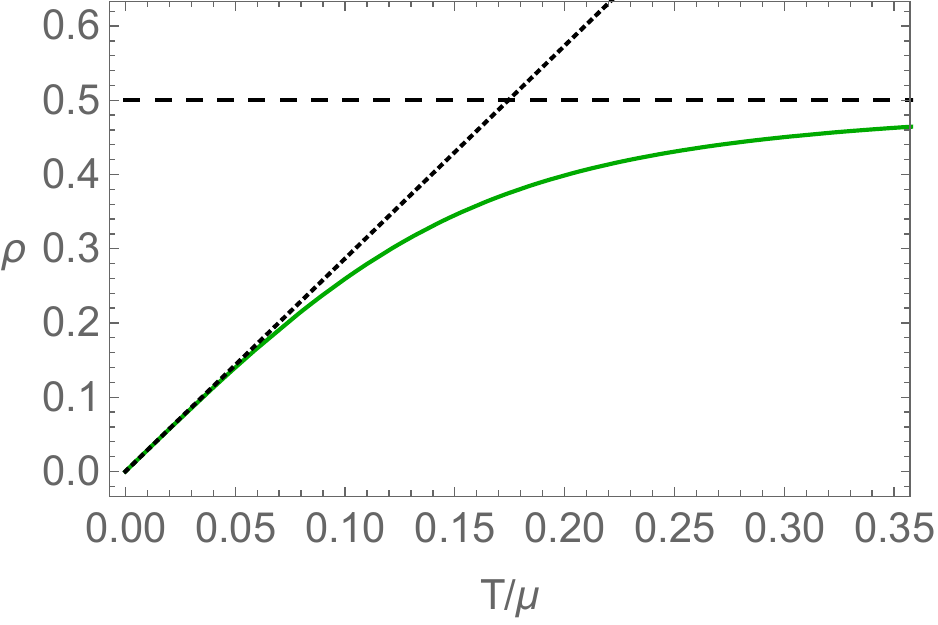} \label{}} 
 \subfigure[$\bb=10$]
     {\ \  \ \includegraphics[width=7cm]{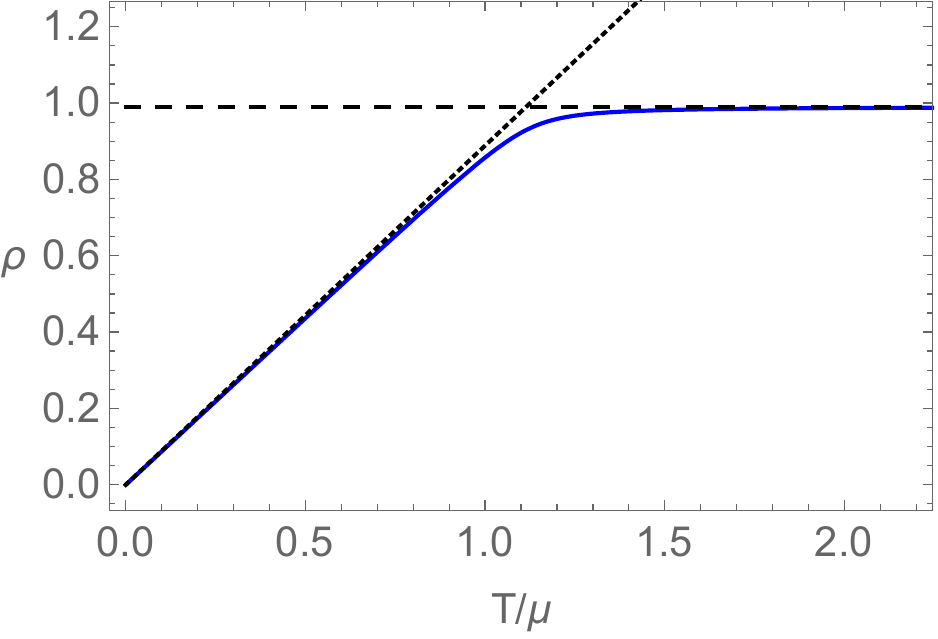} \label{}}
 \subfigure[$\bb=0.1, 1, 10$]
     { \includegraphics[width=7cm]{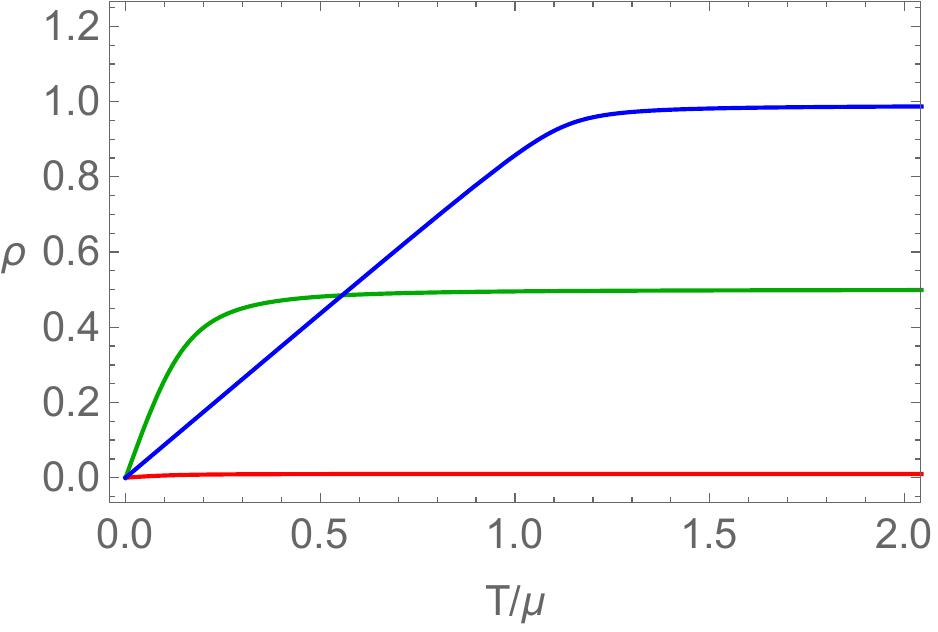} \label{}}
          \caption{ Resistivity vs temperature at fixed $\bb = 0.1, 1, 10$. The horizontal dashed lines are the inverse of \eqref{sigmabigT}. {The dotted lines are \eqref{zaanen1} (a), the inverse of \eqref{sigmasmallT} (b)  and the inverse of \eqref{sigmabigb} (c).}} \label{fig1}
\end{figure}

In \cite{Davison:2013txa}, in order to propose a universal mechanism of the linear-$T$ resistivity, weak momentum relaxation and low temperature limit are considered. It corresponds to the case for both $\bT \ll 1$ and $\bb \ll 1$ from \eqref{sigmasmallT} and \eqref{sigmasmallb}
\begin{equation} \label{zaanen1}
\rho = \frac{1}{\sigma_{DC}} \sim \frac{4\pi \bb^2}{\sqrt{3}}   \bT \,,
\end{equation}
which reproduces $\rho \sim \bb^2 \bT$  in \cite{Davison:2013txa} and here we identify a precise numerical coefficient. 
However, from this asymptotic behavior in the limit $\bT \ll 1$, it is not clear how much linear-$T$ resistivity robust at higher (but still low) temperature\footnote{We will clarify the meaning of `low temperature' later in terms of the superconducting phase transition temperature.}, for example, at $\bT \lesssim 1$. To check it we make an exact plot of the resistivity, the inverse of \eqref{conduct1}, without any approximation in Fig. \ref{fig1}(a), where $\bb = 0.1$. The red curve is the exact resistivity, the horizontal dashed lines are the inverse of \eqref{sigmabigT} and the dotted lines are \eqref{zaanen1}.  We see that the linear-$T$ behavior of resistivity is not so robust at small temperature.

However, from \eqref{sigmabigb} we find that there is another mechanism for linear-$T$ resistivity: strong momentum relaxation. To check it we make exact plots of the resistivity for $\bb=1$ (Fig. \ref{fig1}(b)) and $\bb=10$ (Fig. \ref{fig1}(c)). As $\bb$ increases, linear-$T$ behavior is retained for higher temperature. 
Note that, for $\bb \gtrsim 1$, \eqref{zaanen1} is not a good approximation  at small temperature. Instead, we have to use \eqref{sigmasmallT} or \eqref{sigmabigb}  which correspond to the dotted lines in Fig. \ref{fig1}(b) and (c) respectively.    Fig. \ref{fig1}(d) is presented for comparison for different $\bb$s. 
As shown in \eqref{sigmabigb}, the linear-$T$ comes from the $\tQ \sim 1/\bT^{2}$, \eqref{tQ3}. 
To have a robust linear-$T$ resistivity at higher $\bT$, the temperature dependence of $\tQ$ should be stable for a long range of $\bT$, which is shown as dashed lines in Fig. \ref{fig4}.

There is another way to see \eqref{sigmabigb}. The limit $\bb \gg 1$ corresponds to $\tmu \rightarrow 0$ at fixed $\tb$, so \eqref{tmueq} implies
\begin{equation}
\tb \sim \sqrt{2} (1+\tQ) \,,
\end{equation}
or $\tQ=0$ or $\tQ = -1$. We discard the last two cases because $\tQ=0$ means the dilaton vanishes and $\tQ = -1$ is thermodynamically unstable~\cite{Kim:2017dgz}. Thus, from \eqref{tTeq}, 
\begin{equation}
\frac{\bT}{\bb}  = \frac{\tT}{\tb} \sim  \frac{1}{2\sqrt{2} \, \pi \sqrt{1+\tQ}} \,, 
\end{equation}
and 
\begin{equation}
\sigma_{DC} \sim \sqrt{1+\tQ} \sim \frac{\bb}{2\sqrt{2} \, \pi \, \bT}  \,,
\end{equation}
which agrees to \eqref{sigmabigb}.

The saturation of the resistivity at large temperature has nothing to do with the Mott-Ioffe-Regel limit. 
This large temperature behaviour can be understood by dimensional analysis. 
The dimension of the electric field is $[E] = 2$ and the dimension of a current density in $p$ spacetime dimensions is  $[J] = p -1$. Thus, the dimension of the resistivity $[\rho] = 3 - p$ so the high temperature behaviour of the resistivity is $\rho \sim T^{3-p}$, which agrees with \eqref{sigmabigTp4}. In our case, $p=3$ so the resistivity becomes constant at high temperature.

In summary, the resistivity is in general
\begin{align}
&\rho =\frac{1}{\sqrt{1+\tQ}}\frac{\bb^{2}}{1+\bb^{2}} \,,
\end{align}
and, for large $\bb$, it is well approximated as
\begin{equation}
\begin{split}
&\rho \sim \frac{2\sqrt{2}\pi}{\bb} \bT \qquad \quad \qquad  \left(\bT \lesssim \frac{\bb}{2\sqrt{2}\pi} \right) \,,   \\
&\rho \sim \frac{\bb^2}{(1+\bb^2)} \sim 1 \quad \qquad  \left(\bT \gtrsim \frac{\bb}{2\sqrt{2}\pi} \right) \,.
\end{split}
\end{equation}
These are different from \eqref{zaanen1}, which is valid only at very small $\bT$ in \cite{Davison:2013txa}. {The critical temperature $\bT=\bb/(2\sqrt{2}\pi)$ is determined by an approximate condition $\rho \sim 1$.}

However, note that in strange metal, linear-$T$ behavior is shown up to room temperature. Is the room temperature small or large?
To build a theoretical model for strange metal, we need to quantify how `small' temperature is `small' compared to which quantity.  For this purpose, it will be good to find some intrinsic scale in the model. We will choose our reference scale as the critical temperature ($T_c$),  the superconducting phase transition temperature. In the next section we find the critical temperature and use it to quantify `small'  or `large' temperature.

\subsection{Linear-$T$ resistivity above the critical temperature}
Now let us consider the superconducting phase. Above the critical temperature, there is only one solution with $\Phi=0$ but below the critical temperature, there are two available solutions, one with $\Phi=0$ and the other with $\Phi\ne0$. When there are two solutions, we need to choose one solution with a lower free energy, which is the case with $\Phi \ne 0$. 
In this case, the complex scalar $\Phi$ falls off as 
\begin{equation}
\Phi = \varphi_1 z+\varphi_2 z^2 + \cdots\,,
\end{equation} 
near boundary at $z=0$, if $m^2=-2$.    If we choose $\varphi_1$ as a source $\varphi_2$ corresponds to the condensate of the scalar operator. For spontaneous symmetry breaking we impose the boundary condition $\varphi_1 = 0$. We refer to \cite{Kim:2015dna, Kim:2016hzi} for further technical details on superconductor with momentum relaxation, where a different model so called linear axion model was considered but the method of analysis is similar.

\begin{figure}[]
 \centering
     \subfigure[$q=6,  \bb=1$]
     {\includegraphics[width=4.83cm]{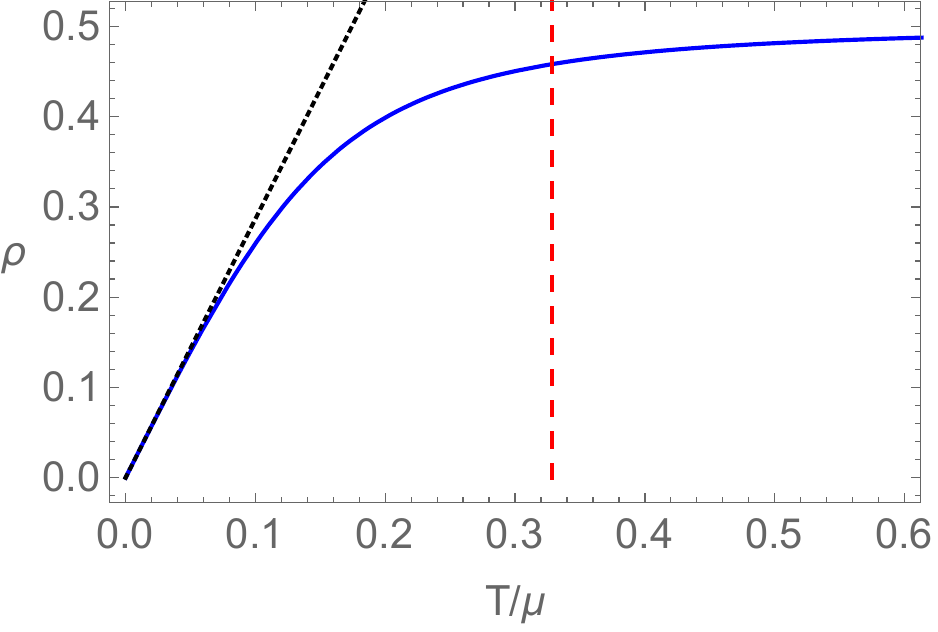} \label{}}
 \subfigure[$q=6,  \bb=3$]
     {\includegraphics[width=4.83cm]{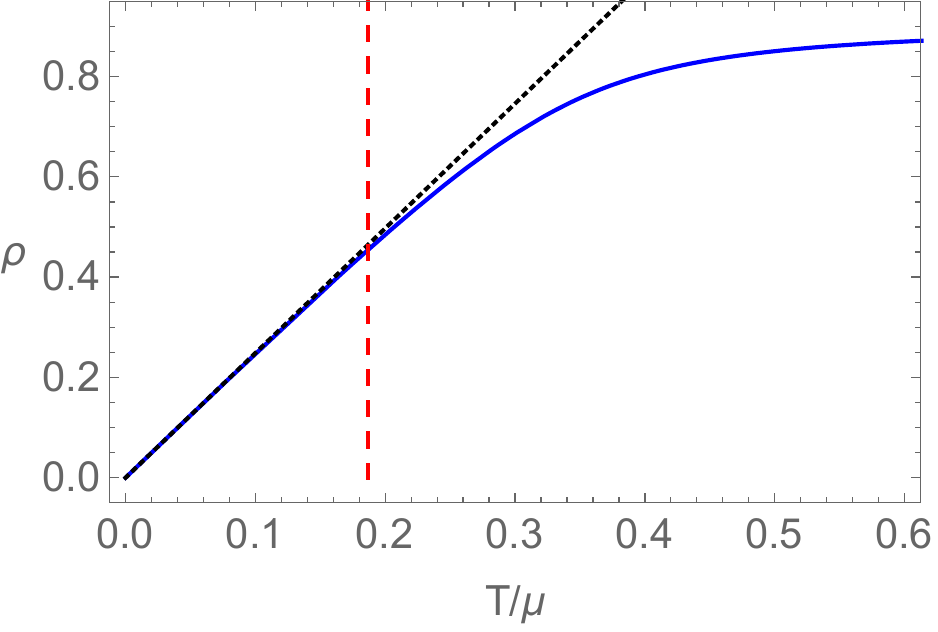} \label{}} 
 \subfigure[$q=6,  \bb=3.3$]
     {\includegraphics[width=4.83cm]{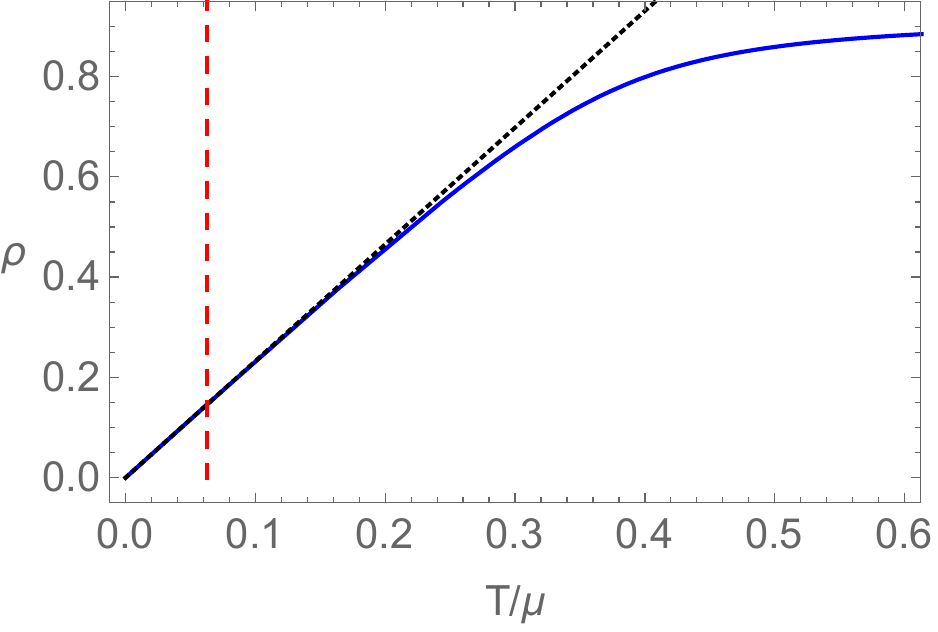} \label{}}
 \centering
     \subfigure[$q=20, \bb=1$]
     {\includegraphics[width=4.83cm]{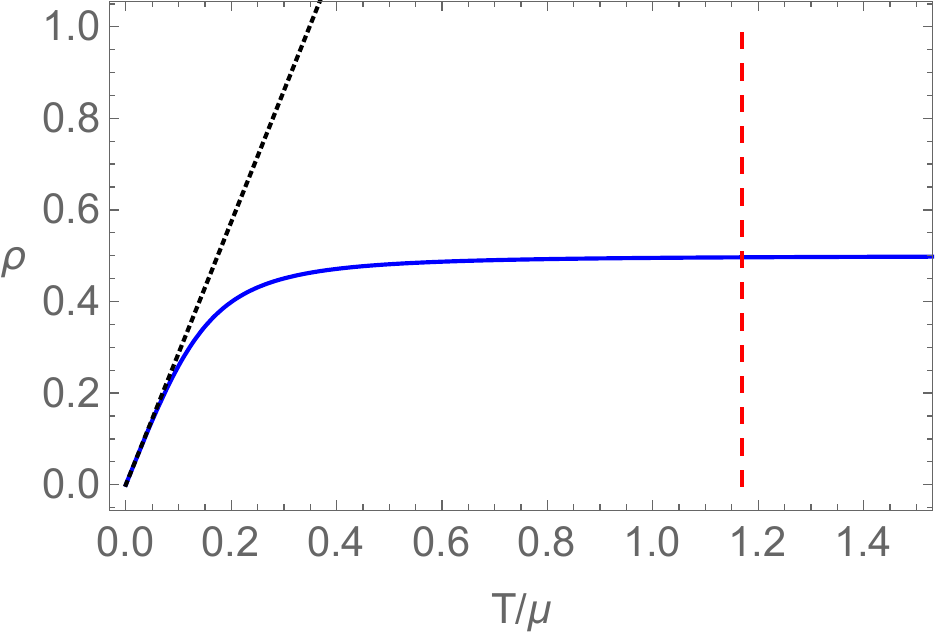} \label{}}
 \subfigure[$q=20, \bb=10$]
     {\includegraphics[width=4.83cm]{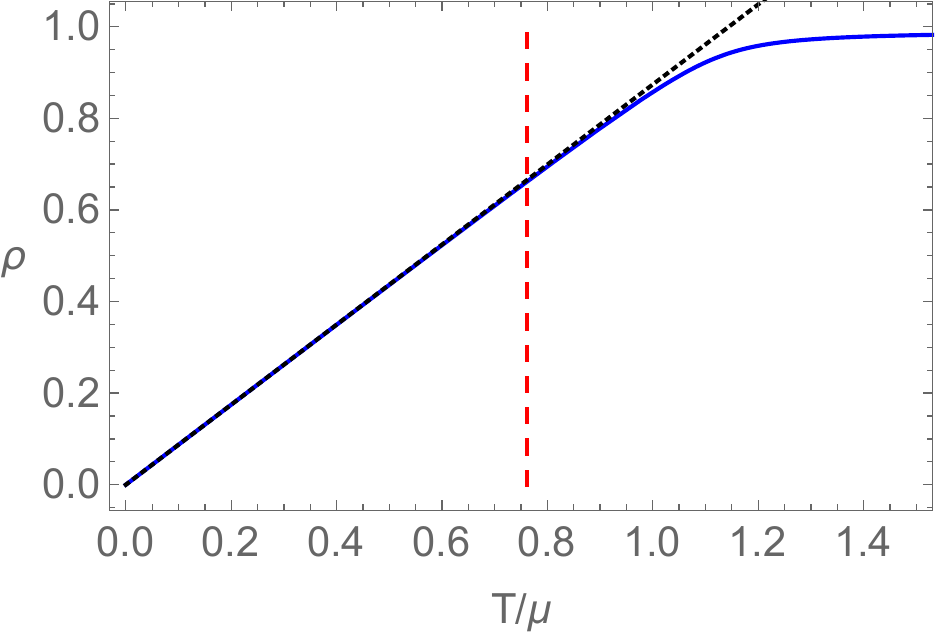} \label{}} 
 \subfigure[$q=20, \bb=11$]
     {\includegraphics[width=4.83cm]{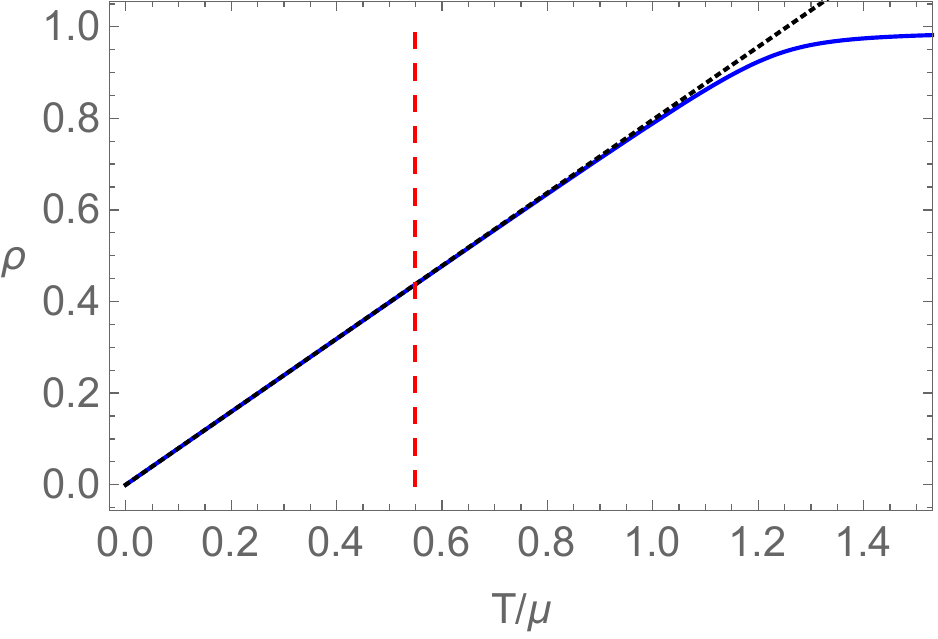} \label{}}
          \caption{ Resistivity vs temperature of the superconductor model with  $q=6$ (top) and $q=10$ (bottom). The vertical red dotted line displays the phase transition temperature, which is higher for bigger $q$ and smaller  $\bb$. For bigger $\bb$, the linear-$T$ resistivity survives above the phase transition as shown in figures (c) and (f).  } \label{fig3}
\end{figure}

The critical temperature depends on $q$ and $\bb$ and we show how the critical temperature is changed by the red vertical dotted lines in  Fig.  \ref{fig3} for $q=6$ and $q=20$. Here $q$ is a charge defined in the covariant derivative in \eqref{action1}.  The critical temperature is lower for smaller $q$ and bigger  $\bb$. 
We use the critical temperature ($T_c$) as our reference and $T>T_c$ ($T<T_c$) is identified with a high (low) temperature.

We find that if the momentum relaxation is strong (large $\bb$), the linear-$T$ resistivity can survive at high temperarue as shown in (c) and (f) of Fig.  \ref{fig3}.  In addition, if  $q$ is small $T_c$ becomes small so it also helps that linear-$T$ resistivity is realized at `high' temperature.


\section{Generalization of the Gubser-Rocha model} \label{sec3}
The advantage of the Gubser-Rocha model studied in section \ref{sec2}  is that the analytic solution is available. Interestingly, it has been shown in \cite{Gouteraux:2014hca} that this model can be generalized in two ways and still allows analytic solutions. First, we may consider arbitrary spacetime dimension. Second, we may have the case in which the IR geometry is not conformal to AdS$_2 \times R^{p-1}$.

Let us  consider a Einstein-Maxwell-Dilaton-Axion action \cite{Gouteraux:2014hca} given by
\begin{equation} \label{GGR}
S=  \int\dd^{p+1}x   \sqrt{-g}   \left[   R -  \frac{1}{2}\partial\phi^2 -  \frac{1}{4} Z(\phi) F^2 + V(\phi) -\frac{1}{2} \sum_{i=1}^{p-1}\partial \psi_{i}^2  \right] \,,
\end{equation}
where the couplings $Z(\phi)$ and $V(\phi)$ are assumed to have the following specific forms:
\begin{align}
\begin{split}
&Z(\phi)=e^{-(p-2)\d\phi} \,, \qquad V(\phi)=V_{1}e^{\frac{\left((p-2)(p-1)\d^{2}-2\right)\phi}{2(p-1)\d}} \,+\, V_{2}e^{\frac{2\phi}{\d-p\d}} \,+\, V_{3}e^{(p-2)\d\phi} \,, \\
\end{split}
\end{align}
with
\begin{align}
\begin{split}
&V_{1}=\frac{8(p-2)(p-1)^3\d^{2}}{\left(2+(p-2)(p-1)\d^2\right)^2}\ , \qquad \quad V_{2}=\frac{(p-2)^{2}(p-1)^2(p(p-1)\d^2-2)\d^{2}}{\left(2+(p-2)(p-1)\d^2\right)^2}\,, \\
&V_{3}=-\frac{2(p-2)^2(p-1)^2\d^2-4p(p-1)}{(2+(p-2)(p-1)\d^2)^2}.
\end{split}
\end{align}

Here $\delta$ is a free parameter. The action \eqref{GGR} is reduced to the Guber-Rocha model in \eqref{action1} if $p=3$ and $\delta = \sqrt{1/3}$.\footnote{{For direct comparison with \eqref{action1} we need to rescale the scalar field as $\phi \rightarrow -\sqrt{3} \, \phi $}}
The equations of motion read 
\begin{align}
\begin{split}
R_{\mu\nu} &= \frac{1}{2}\partial_{\mu}\phi\partial_{\nu}\phi+\frac{1}{2}\sum_{i=1}^{p-1}\partial_{\mu}\psi_{i}\partial_{\nu}\psi_{i}+\frac{Z(\phi)}{2}F_{\mu}{^\rho}F_{\nu\rho}-\frac{Z(\phi)F^2}{4(p-1)}g_{\mu\nu}-\frac{V(\phi)}{p-1}g_{\mu\nu} \,, \\
0&=\grad_{\mu}(Z(\phi)F^{\mu\nu}) \,, \quad 0=\grad^2\psi_{i} \,, \qquad i=1 \dots p-1 \,,\\
0&=\square\phi+V'(\phi)-\frac{1}{4}Z'(\phi)F^2.
\end{split}
\end{align}

With the following ansatz for the solution 
\begin{equation}
\dd s^2=-D(r)\dd t^2+B(r)\dd r^2+C(r)\dd \vec x_{i}^{2} \,, \quad A=A_t(r) \dd t \,, \quad \psi_i=\beta x^i \,,
\end{equation}
the solutions are given by
%
\begin{align}
&\dd s^2=-f(r)h(r)^{\frac{-4}{2+(p-2)(p-1)\d^2}}\dd t^2+h(r)^{\frac{4}{(p-2)(2+(p-2)(p-1)\d^2)}}\left[\frac{\dd r^2}{f(r)}+r^2\dd \Sigma_{p-1}^{2}\right]\ ,  \nonumber \\
&\qquad f(r)=r^2 \left(h(r)^{\frac{4 (p-1)}{(p-2) \left(2+(p-1)(p-2) \delta ^2\right)}}-\frac{r_h^p}{r^{p}} h(r_h)^{\frac{4 (p-1)}{(p-2) \left(2+(p-1)(p-2) \delta ^2\right)}}\right)-\frac{\beta^2 \left(1-\frac{r_h^{p-2}}{r^{p-2} }\right)}{2(p-2)}\,,  \nonumber \\
&\qquad h(r)=1+\frac{Q}{r^{p-2}}\,, \label{sol11}\\
&e^\phi=h(r)^{\frac{-2(p-1)\delta }{2+(p-2)(p-1)\delta ^2}}\ \,, \nonumber \\
&A_t(r)=2\sqrt{(p-1)Q}\frac{\sqrt{(p-2)r_h^{2+p} h(r_h)^{\frac{2 \left(2-(p-2)^2 (p-1) \delta ^2\right)}{(p-2) \left(2+(p-1)(p-2) \delta ^2\right)}}- \frac{r_h^{p}\beta^2}{2 h(r_h)}}}{(p-2)r_h^{p-1} h(r)\sqrt{2+(p-2)(p-1) \delta ^2}}\left(1-\frac{r_h^{p-2}}{r^{p-2}}\right),  \nonumber
\end{align}
where $r_{h}$ is the horizon position. 
The conductivity can be expressed in terms of the horizon data
\begin{equation}\label{genpcon}
\sigma_{DC} := C^{\frac{p-3}{2}}Z+\frac{q^2}{\beta^2YC^{\frac{p-1}{2}}}\Bigr|_{r\rightarrow r_{h}} \,, 
\end{equation}
where $q$ is the charge density
\begin{equation}
q := \frac{A_t' Z}{(BD)^{\frac{1}{2}}C^{\frac{-(p-1)}{2}}}\Bigr|_{r\rightarrow r_{h}} \,. 
\end{equation}

Note that the UV geometry is always asymptotically AdS spacetime. If $\delta =0 $ the solution is simply AdS-RN and the IR geometry is AdS$_2 \times R^{p-1}$. There is a specific value $\delta = \sqrt{\frac{2}{p(p-1)}}$, with which the IR geometry is conformal to AdS$_2 \times R^{p-1}$~\cite{Gouteraux:2014hca}.  After first investigating the former we will consider the latter. 

\subsection{Conformal to AdS$_2 \times R^{p-1}$ IR geomety: $\d$ = $\sqrt{\frac{2}{p(p-1)}}$}
For $p=3$,  \eqref{sol11} is the same as \eqref{ansatz1} with \eqref{sol01} by a coordinate transformation  \eqref{tildes} and footnote 4.  Thus, this case amounts to the high dimensional extension of the Gubser-Rocha model.
The temperature and chemical potential read
\begin{align}
& T = {\frac{1}{4\pi } \left. \frac{|D'|}{\sqrt{D B}}\right|_{r_{h}}} =  r_{h} \, \frac{2p \, (1+\tQ)^{\frac{2}{p-2}} - \tb^2}{8 \pi (1+\tQ)^{\frac{p}{2(p-2)}}} =: r_{h} \, \tT \,, \label{Tp} \, \\
&\mu =  A_t(\infty) = r_{h} \, \sqrt{\frac{p}{p-2} \tQ (1+\tQ)^{\frac{4-p}{p-2}} \left(1 - \frac{\tb^2}{2(p-2)(1+\tQ)^\frac{2}{p-2}} \right) } =: r_{h} \, \tmu \,, \label{mup} 
\end{align}
where 
\begin{equation}
\tQ := \frac{Q}{r_{h}^{p-2}} \,, \quad \tb := \frac{\beta}{r_{h}} \label{betap}\,.
\end{equation}
In order to compute the conductivity as a function of $T$ and $\beta$ at fixed $\mu$, we define
\begin{align}
&\bT := \frac{T}{\mu} = \frac{\tT}{\tmu} =  \frac{2p(p-2)(1+\tQ)^{\frac{2}{p-2}} - (p-2)\tb^2}{4\sqrt{2p} \, \pi \sqrt{\tQ (1+\tQ)^{\frac{2}{p-2}} \left( 2(p-2)(1+\tQ)^{\frac{2}{p-2}} -\tb^2 \right) }  } \,,  \label{bteqp} \\ 
&\bb := \frac{\beta}{\mu} = \frac{\tb}{\tmu} =  \sqrt{\frac{2(p-2)^2 (1+\tQ) \tb^2}{p \, \tQ \left(2(p-2)(1+\tQ)^\frac{2}{p-2} -\tb^2 \right)}} \,, \label{bbeqp}
\end{align}
where we used \eqref{Tp}, \eqref{mup} and \eqref{betap}.

From the formula \eqref{genpcon}, the conductivity is
\begin{align}
\sigma_{DC}  &=  r_{h}^{p-3} \, \left((1+\tQ)^{1+\frac{4-p}{4-2p}} + \frac{(p-2)^2 (1+\tQ)^{\frac{(p-2)^2 - (p-4)}{2(p-2)(p-1)}} }{\bb^2} \right) \,, \label{sigmaDC1}
\end{align}
and the dimensionless conductivity ($\bar{\sigma}_{DC}$)  scaled by the chemical potential \eqref{mup} is defined as
\begin{align}
&\bar{\sigma}_{DC}   :=  \frac{\sigma_{DC}}{\mu^{p-3}}  =  \left( \sqrt{1+\tQ} + \frac{(p-2)^2 \sqrt{1+\tQ}^{\frac{2-(p-3)p}{(p-2)(p-1)}}}{\bb^2} \right) \left(\frac{\bb^2}{2(p-2)} + \frac{(p-2)(1+\tQ)}{p \, \tQ} \right)^{\frac{p-3}{2}} \,, \label{sigmaDCbar} 
\end{align}
Here we used the chemical potential $\mu$ expressed in terms of $\bb$ and $\tQ$:
\begin{align}
\mu  &= r_{h} \, \sqrt{\frac{p}{p-2} \tQ (1+\tQ)^{\frac{4-p}{p-2}} \left(1 - \frac{ p \, \tQ \, \bb^2 }{   2(p-2)^2 (1+\tQ) + p \, \tQ \, \bb^2  }    \right) }.  \label{mubetabar} 
\end{align}
which is obtained from \eqref{mup} by using \eqref{bbeqp}.

The conductivity \eqref{sigmaDCbar} is a function of $\bb$ and $\tQ(\bT, \bb)$. 
Like in section \ref{sec2}, because the analytic expression of $\tQ(\bT, \bb)$ is very complicated and not so illuminating  we only consider its asymptotic form in some limits.
\begin{align}
&\tQ \sim \frac{p \left((p-2)^2 + \bb ^2 \right)^2}{8 \pi^2 (p-2) \left(2(p-2)^{2} +p \, \bb^{2} \right)}\frac{1}{\bT^2}  \,,\,\,\,     \quad             (\bT \ll 1) \,, \label{QsmallT} \\ 
&\tQ \sim \frac{p(p-2)}{16\pi^{2}\bT^2}     \,,\,  \qquad  \qquad \qquad \qquad \qquad \qquad              (\bT \gg 1 \ \ \mathrm{or} \  \ \bb \ll 1) \,,  \label{QlargeT} \\
&\tQ \sim \frac{\bb^2}{8(p-2)\pi^2 \bT^2}  \,,    \qquad \qquad \qquad \qquad \qquad                           (\bb \gg 1) \,.  \label{Qlargeb} 
\end{align}
In the above limits the conductivity \eqref{sigmaDCbar} behaves as 
\begin{equation}
\begin{split}
\bar{\sigma}_{DC}   &\sim    \sqrt{\tQ} \left( 1+\frac{(p-2)^2}{\bb^2}\delta_{p,3} \right)  \left( \frac{p-2}{p} + \frac{\bb^2}{2(p-2)} \right)^\frac{p-3}{2}  \, \qquad \qquad \qquad  \quad   ( \bT \ll 1)  \\
& \sim   \frac{p^2 (p-2) \left( (p-2)^2 + \bb^2 \right) \left( 1-\frac{2}{p} +\frac{\bb^2}{2(p-2)} \right)^\frac{p}{2} }{\pi \left( 2(p-2)^2 + p \, \bb^2 \right)^2} \left( 1+\frac{(p-2)^2}{\bb^2}\delta_{p,3} \right) \frac{1}{\bT}\,,  \label{sigmasmallTp4}  \\
\end{split}
\end{equation}
\begin{align}
\begin{split}
\bar{\sigma}_{DC}   &\sim   \left(1+\frac{(p-2)^2}{\bb^2}\right) \left(\frac{p-2}{p \, \tQ}\right)^\frac{p-3}{2}    \sim \left(1 + \frac{(p-2)^2}{\bb^2}\right) \left(\frac{4\pi}{p} \right)^{p-3} \bT^{p-3}\,, \ \   (\bT \gg 1)  \label{sigmabigTp4}     \\
\end{split}
\end{align}
for given $\bb$ and
\begin{align}
\begin{split}
\qquad\qquad\qquad  \bar{\sigma}_{DC}   &\sim   \frac{1}{\bb^2} \frac{(p-2)^\frac{p+1}{2} (1+\tQ)^{\frac{(p-7)p^2+2(7p-2)}{2(p-2)(p-1)}}  }{(p \, \tQ)^{\frac{p-3}{2}}} \\
& \sim  \frac{(4 \pi)^{p+1}}{\bb^2}  \frac{ (p-2)^2 \left( 1+\frac{(p-2)p}{16\pi^2 \bT^2} \right)^{\frac{4+(p-2)(p-1)p}{2(p-2)(p-1)}} }{p^{p-3} \left( (p-2)p +16\pi^2 \bT^2 \right)^2} \bT^{p+1}    \,, \  \  (\bb \ll 1)  \label{sigmasmallbp4} 
\end{split}
\end{align}
\begin{align}
\begin{split}
\qquad\qquad\qquad\quad  \bar{\sigma}_{DC}   \sim  \sqrt{\frac{\bb^{2(p-3)}}{2^{p-3}(p-2)^{p-3}}} \sqrt{\tQ}  \sim  \sqrt{\frac{\bb^{2(p-2)}}{2^{p}(p-2)^{p-2}}}\frac{1}{\pi \, \bT} \,,   \  \   (\bb \gg 1 )   \label{sigmabigbp4}
\end{split}
\end{align}
for given $\bT$.
 
There are two ways to obatin the linear-$T$ resistivity($\rho=1/\bar{\sigma}_{DC}$) regardless of the dimension $p$: $\bT \ll 1$\, \eqref{sigmasmallTp4} and  $\bb \gg 1$\, \eqref{sigmabigbp4}. The latter (strong momentum relaxation) is robust for a large range of temperature, which is the property we want.  To trace the origin of this linear-$T$ behavior, let us revisit the structure of the conductivity ($\bar{\sigma}_{DC}$) defined in \eqref{sigmaDCbar}. 
Without simplifying the expression as in  \eqref{sigmaDCbar}, the conductivity reads
\begin{align}
\begin{split}
\bar{\sigma}_{DC}  :=  \frac{\sigma_{DC}}{\mu^{p-3}}  = \frac{ \left((1+\tQ)^{1+\frac{4-p}{4-2p}} + \frac{(p-2)^2 (1+\tQ)^{\frac{(p-2)^2 - (p-4)}{2(p-2)(p-1)}} }{\bb^2} \right)  }{ \sqrt{\frac{p}{p-2} \tQ (1+\tQ)^{\frac{4-p}{p-2}} \left(1 - \frac{ p \, \tQ \, \bb^2 }{   2(p-2)^2 (1+\tQ) + p \, \tQ \, \bb^2  }    \right) }^{ \,p-3}   } \label{sigmaDCbar2}
\end{split}
\end{align}
with \eqref{sigmaDC1} and \eqref{mubetabar}. 
First, in the numerator ($\sigma_{DC}$), the first term is always dominant in the limit $\bb \gg 1$. Note that even though the second term in the numerator is suppressed by $\bb^2$ we also have to check $\bb$ dependence in $\tQ$, \eqref{Qlargeb},  to see if the first term is indeed dominant. Next, the parenthesis in the denominator is simplified as $ {2(p-2)^2}/(p \,\bb^2)$ in the limit $\bb \gg 1$. Thus, 
%
\begin{align}
\bar{\sigma}_{DC} 
 \sim \frac{  \tQ^{1+\frac{4-p}{4-2p}}  }{ \sqrt{\frac{p}{p-2} \tQ \, \tQ^{\frac{4-p}{p-2}} \left(\frac{2(p-2)^2}{p \,\bb^2} \right) }^{ \,p-3} }  
 =  \sqrt{\frac{\bb^{2(p-3)}}{2^{p-3}(p-2)^{p-3}}} \sqrt{\tQ} \,,
\end{align}
which agrees with  \eqref{sigmabigbp4}. In short, the origin of the linear-$T$ dependence of $\bar{\sigma}_{DC}$ for strong momentum relaxation comes from the combination of $\sigma_{DC}$ \eqref{sigmaDC1} and $\mu$ \eqref{mubetabar}.

\begin{figure}[]
 \centering
     \subfigure[ $p=4$, $\bb=1$]
     {\includegraphics[width=4.83cm]{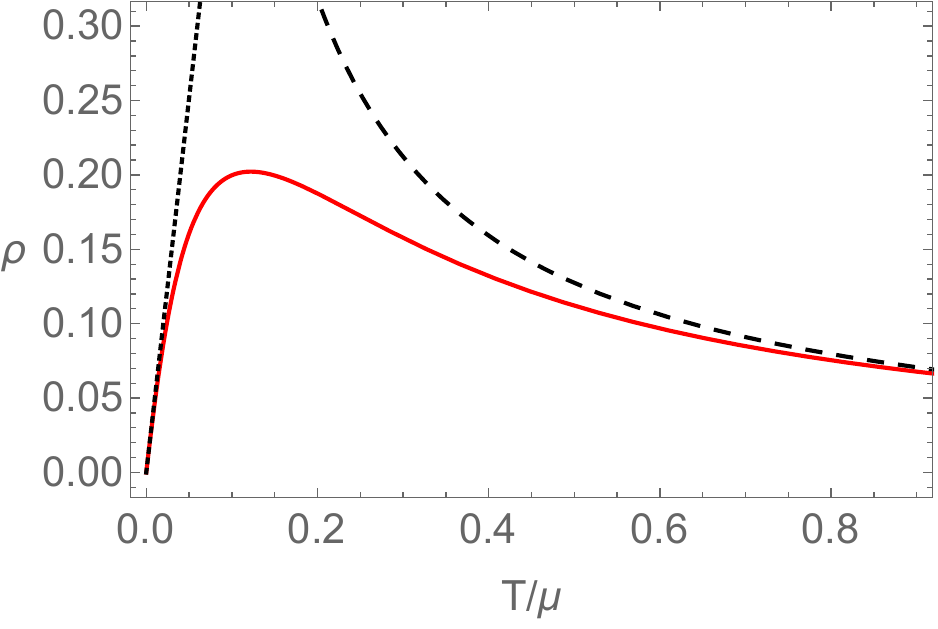} \label{}}
 \subfigure[$p=4$, $\bb=10$]
     {\includegraphics[width=4.83cm]{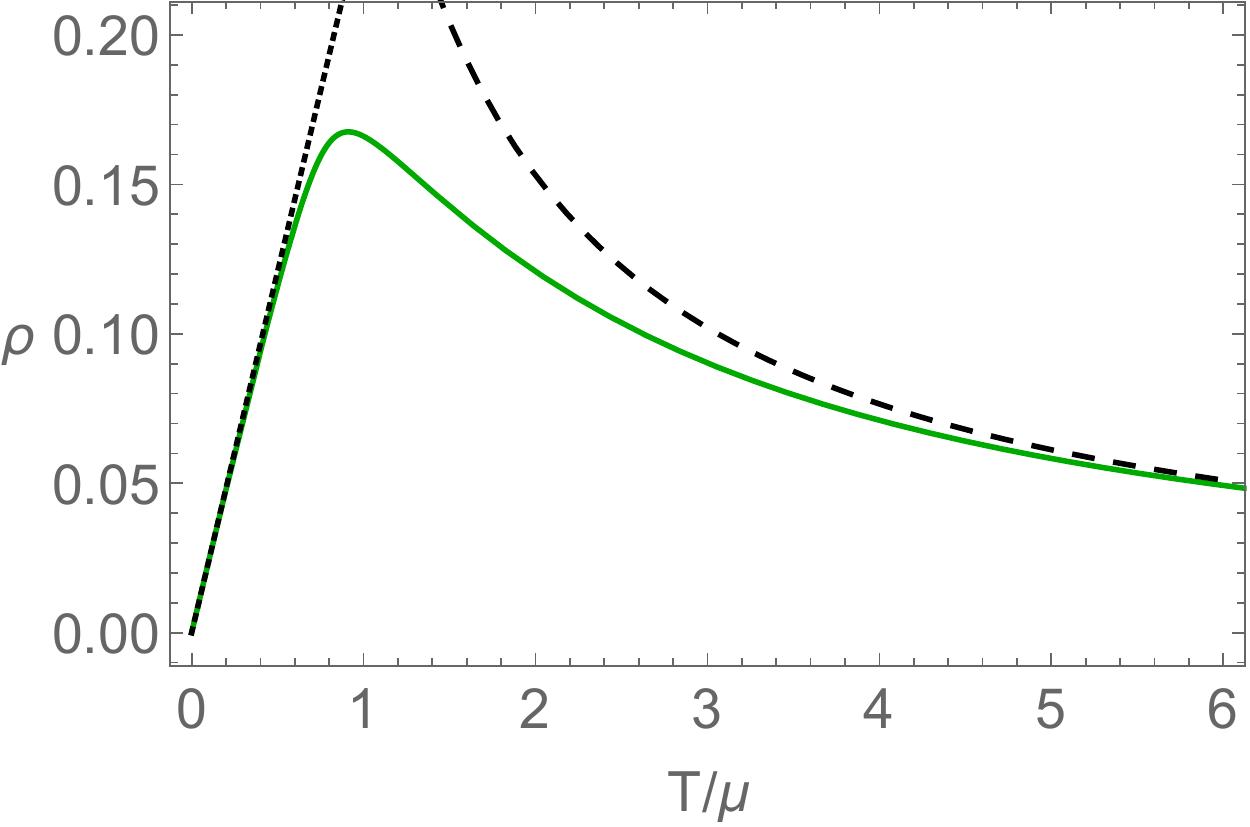} \label{}} 
 \subfigure[$p=4$, $\bb=20$]
     {\includegraphics[width=4.83cm]{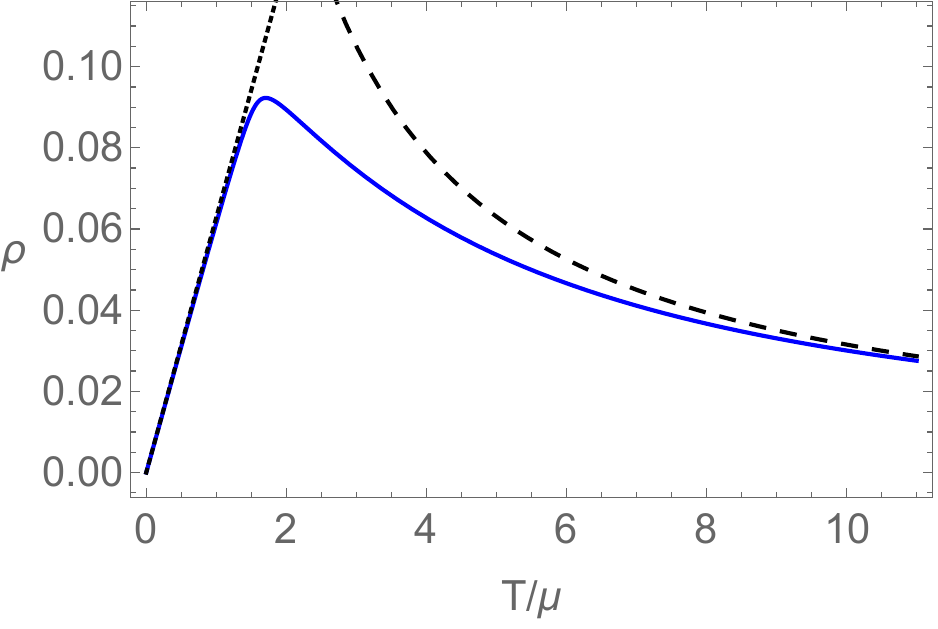} \label{}}
 \subfigure[$p=5$, $\bb=1$]
     {\includegraphics[width=4.83cm]{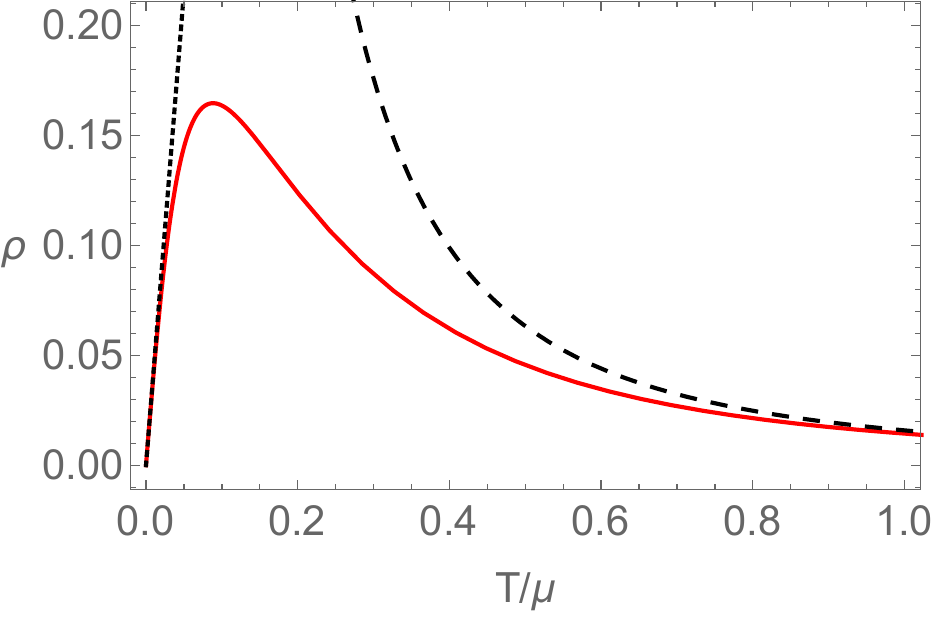} \label{}}
 \subfigure[$p=5$, $\bb=10$]
     {\includegraphics[width=4.83cm]{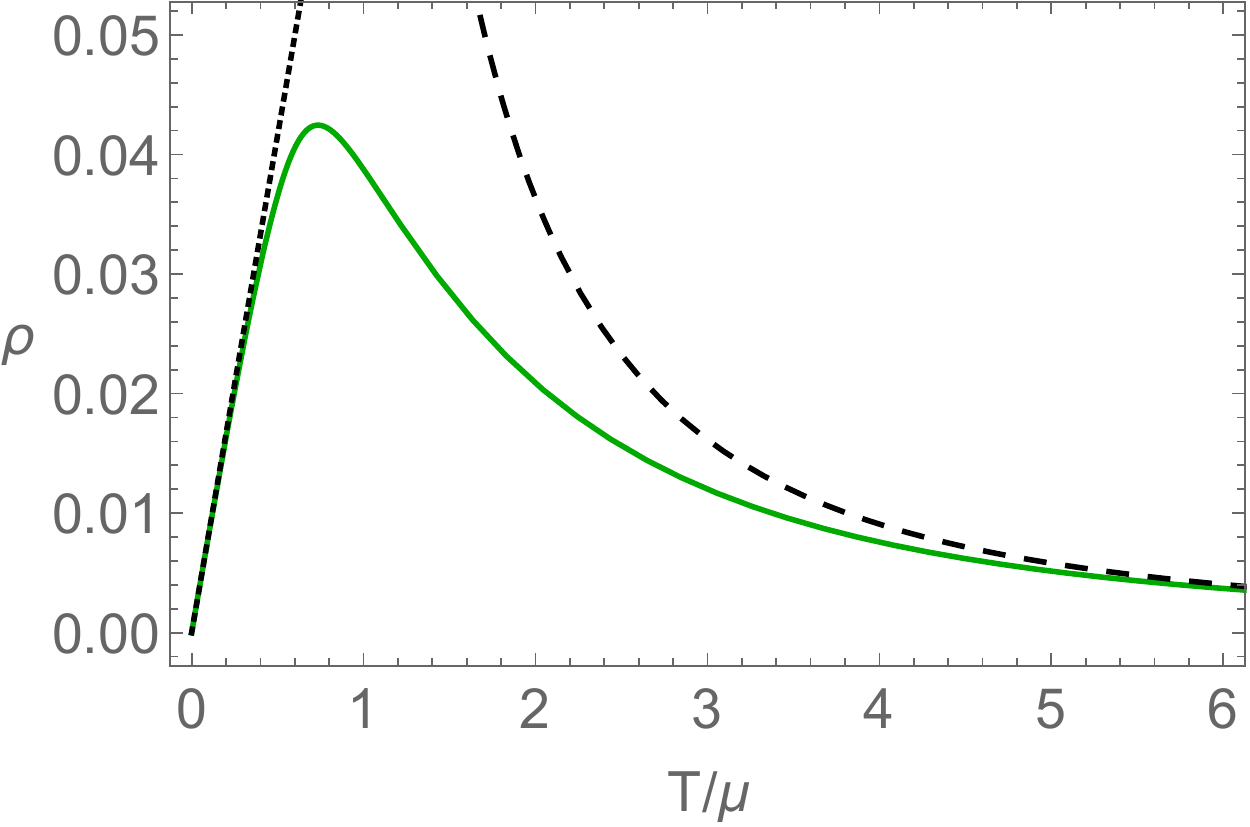} \label{}} 
 \subfigure[$p=5$, $\bb=20$]
     {\includegraphics[width=4.83cm]{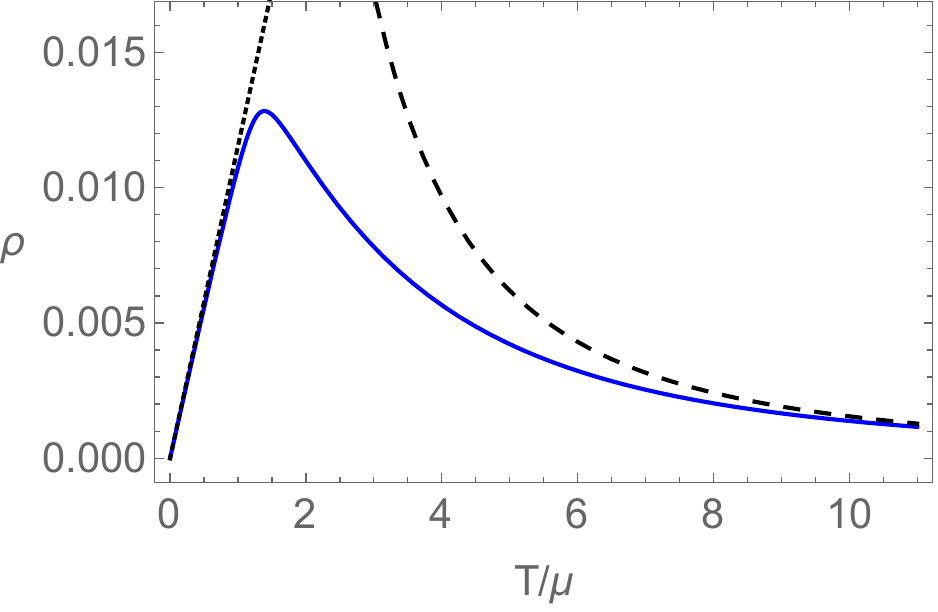} \label{}}
          \caption{Resistivity vs temperature at fixed $\bb = 1, 10, 20$ for $p=4$ (top) and $p=5$ (bottom).      
          The dashed curves (right side) are the inverse of \eqref{sigmabigTp4}. {The dotted lines (left side) are the inverse of \eqref{sigmasmallTp4} for $\bb = 1,10$ and the inverse of \eqref{sigmabigbp4} for $\bb=20$.}} \label{fig5}
\end{figure}

The exact plots of the resistivity \eqref{sigmaDCbar} are shown in Fig. \ref{fig5} for $p=4, 5$ and $\bb =1, 10, 20$. As expected, the large $\bb$ gives a robust linear-$T$ resistivity.  The slop at small temperature agrees with \eqref{sigmasmallTp4} (or \eqref{sigmabigbp4} for $\bb =10, 20$).  At large temperature, the resistivity decreases as $1/T (p=4)$ and $1/T^2 (p=5)$ respectively as shown in \eqref{sigmabigTp4}.

\subsection{AdS$_2 \times R^{p-1}$ IR geomety: $0 \le \d < \sqrt{\frac{2}{p(p-1)}}$}
Let us turn to the case with an arbitrary $\d \in \left[0 , \sqrt{\frac{2}{p(p-1)} } \right) $, where $\delta=0$ corresponds to the AdS-RN balck hole and $\delta = \sqrt{\frac{2}{p(p-1)} } $ corresponds to the Gubser-Rocca model in higher dimensions. 
The temperature and chemical potential read
\begin{align}
T &=  r_{h} \, \frac{2 \left(p(1+\tQ)-\frac{4(p-1)\tQ}{2+(p-2)(p-1)\delta^2}\right)  (1+\tQ)^{\frac{2p-(p-2)^2(p-1)\delta^2}{(p-2)(2+(p-2)(p-1)\delta^2)}} - \tb^2  }{8 \pi (1+\tQ)^{\frac{2(p-1)}{(p-2)(2+(p-2)(p-1)\delta^2)}}    } =: r_{h} \, \tT \,, \, \label{Tpfree}  \\
\mu &= r_{h} \, \sqrt{\frac{4(p-1) \, \tQ (1+\tQ)^{\frac{4-2(p-2)^2(p-1)\delta^2}{(p-2)(2+(p-2)(p-1)\delta^2)}}}{(p-2)(2+(p-2)(p-1)\delta^2)} \left(1 - \frac{\tb^2}{2(p-2)(1+\tQ)^\frac{2p-(p-2)^2(p-1)\delta^2}{(p-2)(2+(p-2)(p-1)\delta^2)}} \right) } \nonumber \\ 
& =: r_{h} \, \tmu \, . \label{mupfree}  
\end{align}

Similarly to \eqref{bteqp} and \eqref{bbeqp} we can define $\bT$ and $\bb$, of which analytic expression is complicated and we do not present here. 
%
The  conductivity \eqref{genpcon} reads
\begin{align}
\sigma_{DC}  &=  r_{h}^{p-3} \, \left((1+\tQ)^{\frac{2(p-3+(p-2)^2 (p-1) \delta^2)}{(p-2)(2+(p-2)(p-1)\delta^2)}} + \frac{(p-2)^2 (1+\tQ)^{\frac{2(3-p+(p-2)^2 (p-1) \delta^2)}{(p-2)(2+(p-2)(p-1)\delta^2) }} }{\bb^2} \right) \,,
\end{align}
and the dimensionless conductivity is defined by
\begin{align} \label{rho123}
\bar{\sigma}_{DC}   :=  \frac{\sigma_{DC}}{\mu^{p-3}}  =  &\left( \sqrt{1+\tQ}^{\frac{(p-2)(p-1)(p+1)\delta^2 -2(p-3)}{2+(p-2)(p-1)\delta^2}}  + \frac{(p-2)^2 \sqrt{1+\tQ}^{\frac{(p-2)^2 (p^2-1)\delta^2 -2(p-3)(p+2)}{(p-2)(2+(p-2)(p-1)\delta^2)}}}{\bb^2} \right)  \nn \\
  &\times \left(\frac{\bb^2}{2(p-2)} + \frac{(p-2)(1+\tQ)(2+(p-2)(p-1)\delta^2)}{4(p-1) \, \tQ} \right)^{\frac{p-3}{2}}. 
\end{align}
%

\begin{figure}[]
\centering
     \subfigure[$\bb=0.1$]
     {\includegraphics[width=7.44cm]{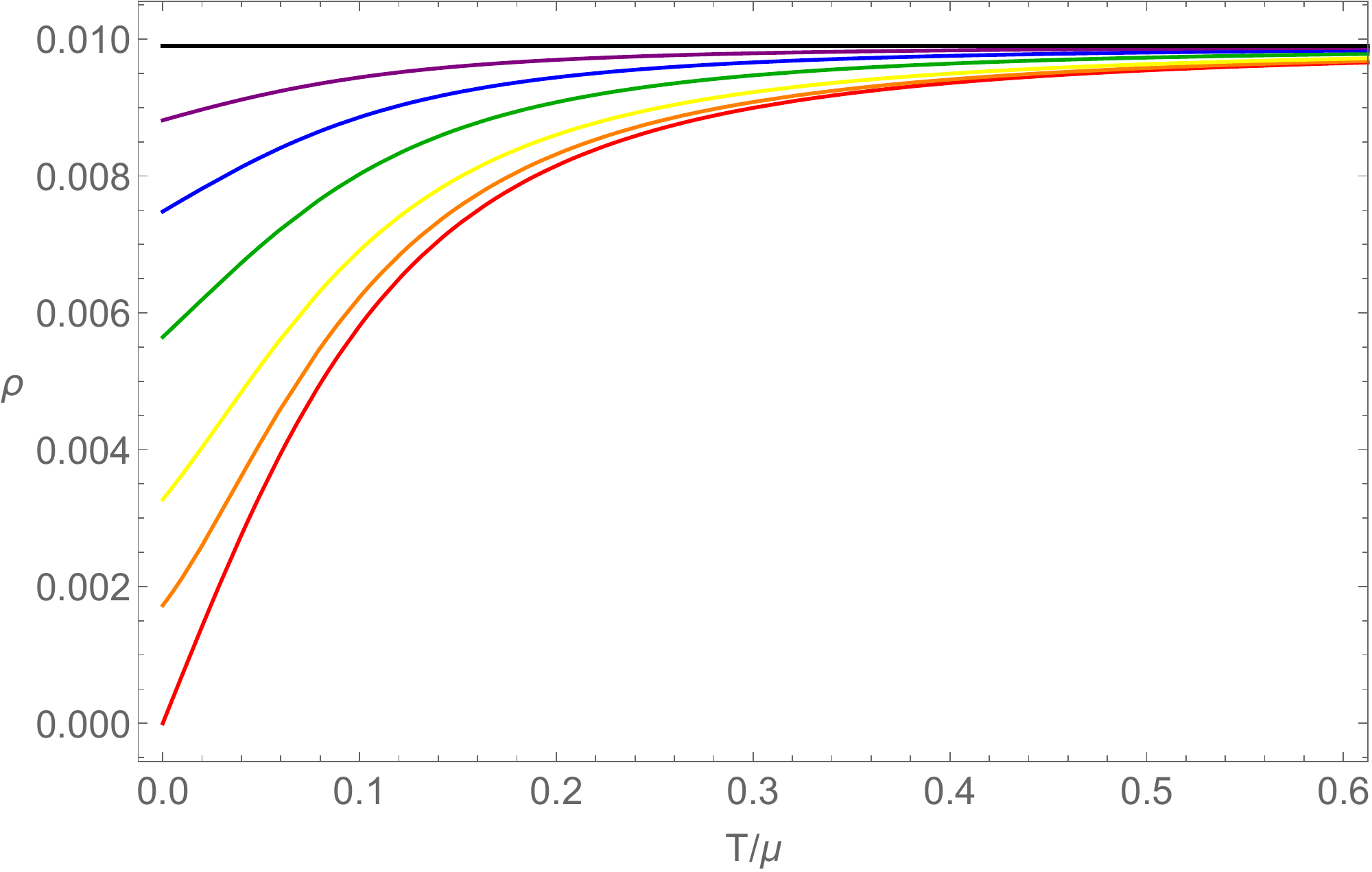} \label{}}
     \subfigure[$\bb=10$]
     {\includegraphics[width=7.25cm]{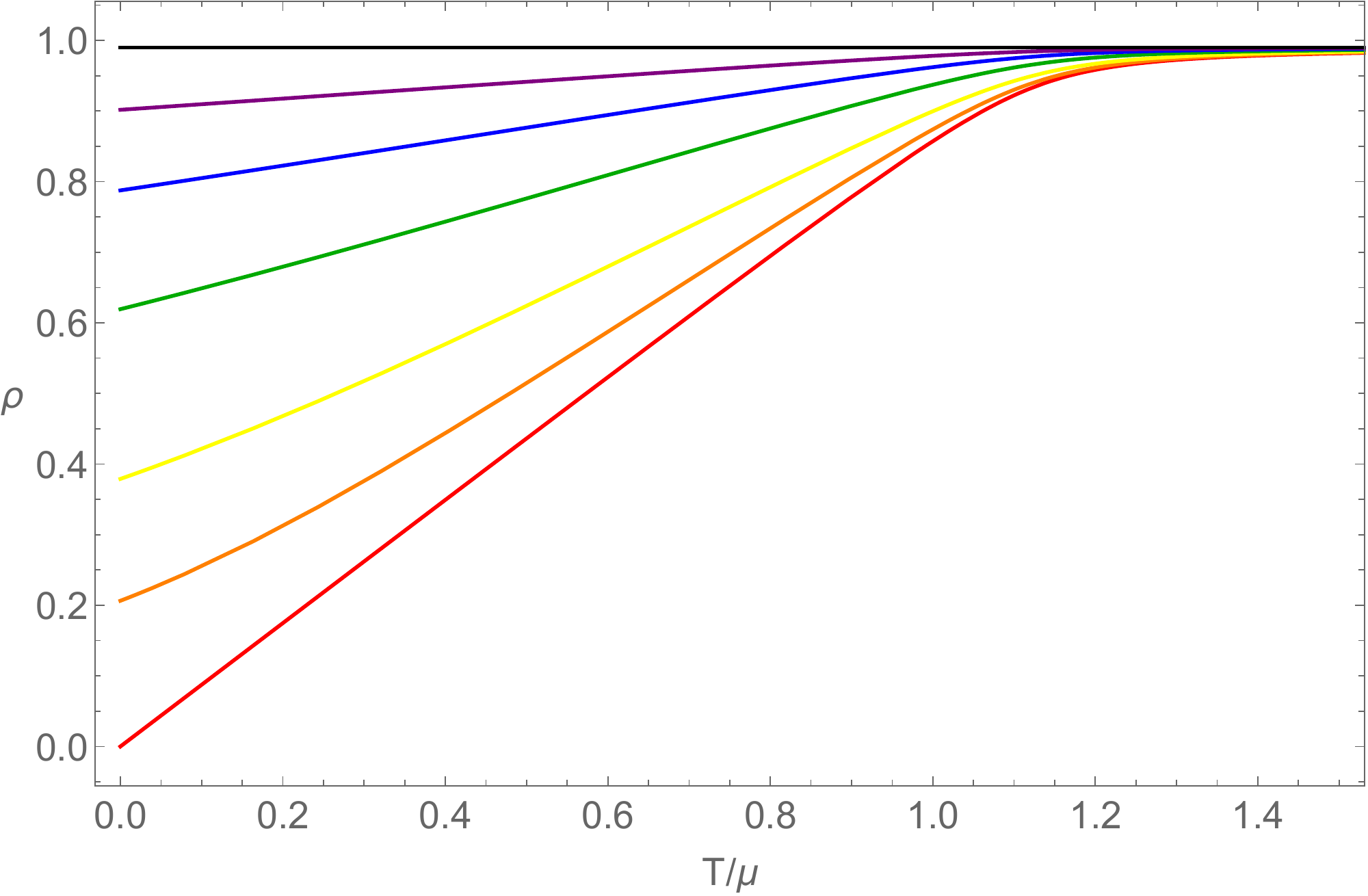} \label{}} 
 \caption{Resistivity ($\rho$) vs temperature.  $\bb = 0.1$(a), 10(b) for $p=3$.  Various colors represent different $\d$s.  i.e.  $\d$ = $\frac{1}{\sqrt{3}}, \frac{55}{100}, \frac{50}{100}, \frac{40}{100},\frac{30}{100},\frac{20}{100},0$ (red, orange, yellow, green, blue, purple, black). }\label{fig6}
\end{figure}

We make a plot of the resistivity (1/$\bar{\sigma}_{DC}$) for several $\d$s in Fig. \ref{fig6}.  Also in this case, we can see a qualitative tendency that the strong momentum relaxation gives a more robust linear-$T$ resistivity at higher temperature up to residual resistivity at zero temperature. 
The residual resistivity at zero temperature is 
\begin{align}
\rho  \sim \frac{\bb^2}{1+\bb^2}\left( \frac{(1+3\d^2 +2\bb^2)-\sqrt{\left(1+3\d^2+2\bb^2\right)^2 - (3\d^2-1)(3+3\d^2+6\bb^2)}}{2\bb^2-2-\sqrt{4+(10-6\d^2)\bb^2+4\bb^4}} \right)^{\frac{2\d^2}{1+\d^2}} \,,
\end{align}
for $p=3$.
This can be obtained by plugging $\tQ(\d, \bb)$  in \eqref{rho123}, where $\tQ$ is computed by requiring  $T=0$ in \eqref{Tpfree}. The expression is simplifed in two limits:
\begin{align}
\rho  &\sim  \left( \frac{1-3\d^2}{4} \right)^{\frac{2\d^2}{1+\d^2}} \, \bb^2  \,, \qquad           (\bb \ll 1) \,,  \\
\rho  &\sim  \left( \frac{1-3\d^2}{3-\d^2} \right)^{\frac{2\d^2}{1+\d^2}}  \,. \qquad\quad\,\,   (\bb \gg 1)
\end{align}
In strong momentum relaxation, the residual resistivity is independent of $\bb$, and is  only a function of $\d$.
Only for $\delta = \sqrt{\frac{1}{3} }$ corresponding to the Gubser-Rocha model, $\rho=0$ at zero temperature. For $\d=0$ corresponding to the AdS-RN black hole case, $\rho=\bb^2/(1+\bb^2)$ at  all temperature.


\section{Conclusion} \label{sec4}

In this paper, we have studied resistivity in extended classes of the Gubser-Rocha model with momentum relaxation. The IR geometry of these models  is AdS$_2 \times R^{p-1}$ or conformal to  AdS$_2 \times R^{p-1}$.  For the former, there is a residual resistivity at zero temperature, while for the latter, the resistivity vanishes at zero temperature.

Investigating the linear-$T$ resistivity at higher temperature is important because linear-$T$ resistivity is observed even in room temperature well above the superconducting phase transition temperature (critical temperature).  However, most holographic studies have been focused on the `low' temperature limit.  To our knowledge, our work is the first holographic study considering linear-$T$ resistivity at higher temperature regime above the critical temperature. We have shown that, in  the Gubser-Rocha model and its several variants,  if momentum relaxation becomes strong enough, the linear-$T$ resistivity in holographic models becomes  more robust up to  higher temperature and is realized above the critical temperature. Our result is also contrast to the well known result in~\cite{Davison:2013txa} where week momentum relaxation is essential. 

To see how much this observation is universal, it will be important to investigate other holographic models such as scaling geometries in~\cite{Gouteraux:2014hca,Ahn:2017kvc}.  In these cases, only the solutions at low temperature limit were known analytically and the critical exponents for linear-$T$ resistivity have been specified.  For finite temperature regime, we should resort to numerical solutions with UV completing potentials. It seems that strong momentum relaxation plays an important role for linear-$T$ resistivity also in these models~\cite{WIP}. 

Our model is a particularly interesting model to investigate the Homes' law~\cite{Blauvelt:2017koq,Homes:2004wv}. Homes' law is a universal relation between  superfluid density at zero temperature $\rho_s(T=0)$, critical temperature ($T_c$), and DC electric conductivity right above the critical temperature ($\sigma_{\mathrm{DC}}(T_c)$):
\begin{equation}
\rho_s(T=0) =  C \sigma_{\mathrm{DC}}(T_c) T_c\,,
\end{equation}
where $C$ is a material independent universal number.
There have been some works to understand the Home's law from holographic perspective~\cite{Erdmenger:2015qqa,Kim:2015dna,Kim:2016hzi,Kim:2016jjk}. In those works Homes' law was observed within some parameter windows, but more fundamental understanding is still lacking. Most superconducting materials exhibiting Homes' law also show linear-$T$ resistivity in normal phase. Indeed, the linear-$T$ resistivity was proposed to play a fundamental role in Homes' law~\cite{Homes:2004wv}. Because our holographic model turns out to have linear-$T$ resistivity in normal phase, contrary to the models in \cite{Erdmenger:2015qqa,Kim:2015dna,Kim:2016hzi,Kim:2016jjk}, it will be a proper set-up to study Homes' law. 

Another interesting property we may investigate in our model is a conjectured universal lower bound
\begin{equation}
C_e := \frac{2 \pi T D}{v_B^2} \,,
\end{equation}
where $D$ is a heat diffusion constant, $T$ is temperature,  $v_B$ is the butterfly velocity~\cite{Hartnoll:2014lpa, Blake:2016wvh, Blake:2016jnn, Blake:2017qgd, Kim:2017dgz, Ahn:2017kvc}.  At zero temperature limit in strong momentum relaxation regime, for the case with IR geometry conformal to AdS$_2 \times R^{p-1}$, $C_e$ is $1$~\cite{Kim:2017dgz} and for the case  AdS$_2 \times R^{p-1}$ case, $C_e$ is expected between $1/2$ and 1 as shown in \cite{Blake:2017qgd}. However, this analysis is valid only at zero temperature limit, so how much it is robust at high temperature is not clear.  Given that shear viscosity to entropy density ratio KSS (Kovtun-Son-Starinets) bound is robust at high temperature, it will be interesting to see if $C_e$ is also robust at high temperature.


\acknowledgments

We would like to thank Yongjun Ahn for valuable discussions and correspondence.  This work was supported by Basic Science Research Program through the National Research Foundation of Korea(NRF) funded by the Ministry of Science, ICT $\&$ Future Planning(NRF- 2017R1A2B4004810) and GIST Research Institute(GRI) grant funded by the GIST in 2018. 
We also would like to thank the APCTP(Asia-Pacific Center for Theoretical Physics) focus program,``Geometry and Holography for Quantum Criticality'' in Pohang, Korea for the hospitality during our visit, where part of this work was done.

\appendix

\section{Resistivity in terms of $\mu/\beta$ and $T/\beta$.}
In this appendix, we compute  the resistivity at fixed $\beta$, i.e. $\sigma_{DC} =\sigma_{DC} (\mu/\beta, T/\beta )$. For simplicity, only $p=3$ has been considered. Let us first define
\begin{align}
&\hT    := \frac{T}{\beta} =   \frac{\bT}{\bb}   \,, \qquad   \hm := \frac{\mu}{\beta} = \frac{1}{\bb} \,,
\end{align}
which can be obtained by using \eqref{bteq} and \eqref{bbeq}. The conductivity \eqref{conduct1} reads 
\begin{equation} \label{conductbetafix}
\sigma_{DC} =  \sqrt{1+\tQ}\left(1+ \hm^{2}\right)  \,,
\end{equation}
where $\tQ$ is a function of $\hT$ and $\hm$ and its asymptotic forms are
\begin{align}
&\tQ \sim \frac{3(1+\hm^2)^{2}}{8\pi^{2}(2\hm^2+3)\hT^2}  \,, \qquad (\hT \ll 1) \,, \\ 
&\tQ \sim \frac{3 \, \hm^2}{16\pi^{2}\hT^2}  \,, \, \ \quad  \qquad \qquad (\hT \gg 1 \ \ \mathrm{or} \  \ \hm \gg 1) \,,  \\
&\tQ \sim \frac{1}{8\pi^2 \hT^2}-1 \,, \qquad \qquad \,\, (\hm \ll 1) \,.
\end{align}

In the above limits the conductivity \eqref{conductbetafix} behaves as
\begin{align}
&\sigma_{DC} \sim \sqrt{\tQ}\left(1+\hm^2 \right)  \sim \frac{\sqrt{3}\left(1+\hm^2\right)^2}{2 \pi \sqrt{4\hm^2 + 6}} \frac{1}{\hT}  \,, \qquad\qquad   (\hT \ll 1) \,, \label{sigmasmallTbeta} \\
&\sigma_{DC} \sim 1+\hm^2  \,, \qquad \qquad \qquad \qquad \qquad \qquad \qquad \,\, \ \  (\hT \gg 1) \,  \label{sigmabigTbeta}
\end{align}
 for given $\hm$ and 
\begin{align}
& \sigma_{DC} \sim \sqrt{1+\tQ} \sim   \frac{1}{2\sqrt{2} \, \pi \, \hT}  \,, \qquad\qquad \qquad \qquad \quad\,   (\hm \ll 1) \,, \label{sigmasmallbbeta} \\
&\sigma_{DC} \sim  \sqrt{\tQ} \, \hm^2 \sim  \frac{\sqrt{3} \, \hm^3}{4\pi \hT}  \,,  \qquad\qquad\qquad\qquad\qquad\,   \ \, (\hm \gg 1) \, \label{sigmabigbbeta} 
\end{align}
for given $\hT$. There are three limits showing linear-$T$ resistivity: $\hT \ll 1$, $\hm \ll 1$ and $\hm \gg 1$. 

\begin{figure}[]
 \centering
     \subfigure[$\hm=0$]
     {\includegraphics[width=4.6cm]{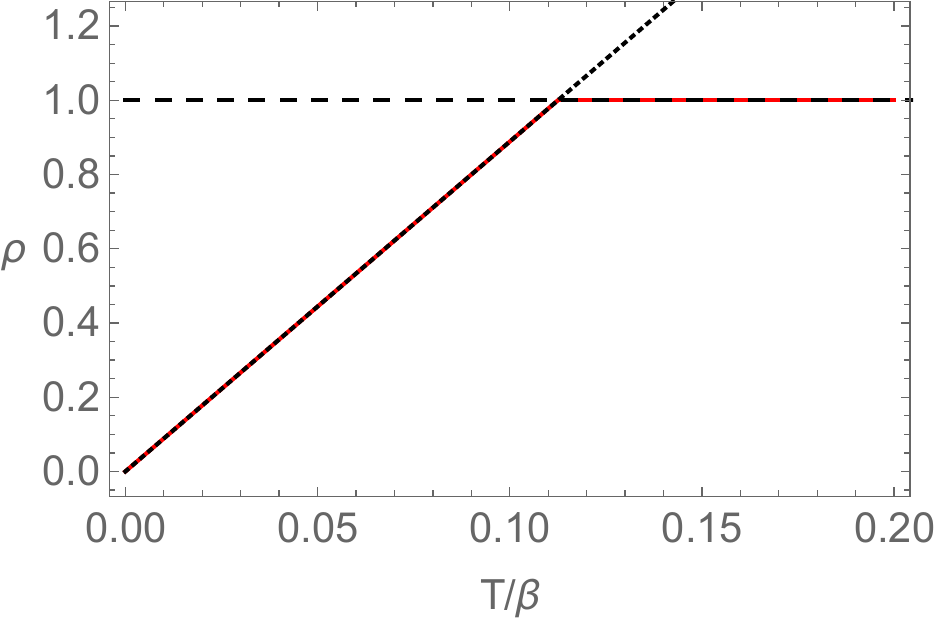} \label{}}
 \subfigure[$\hm=0.1$]
     {\ \  \ \includegraphics[width=4.6cm]{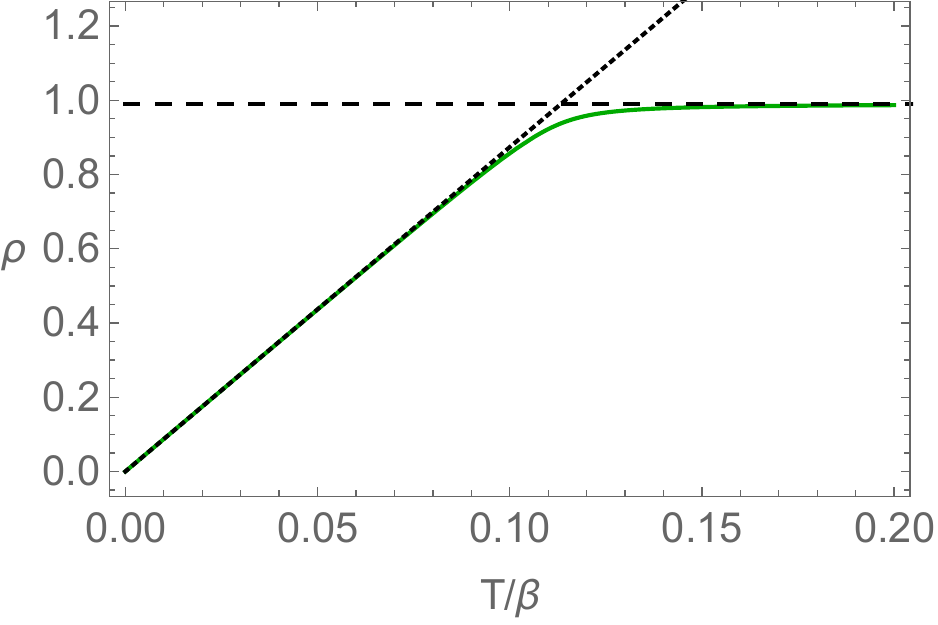} \label{}}
 \subfigure[$\hm=1$]
     { \includegraphics[width=4.6cm]{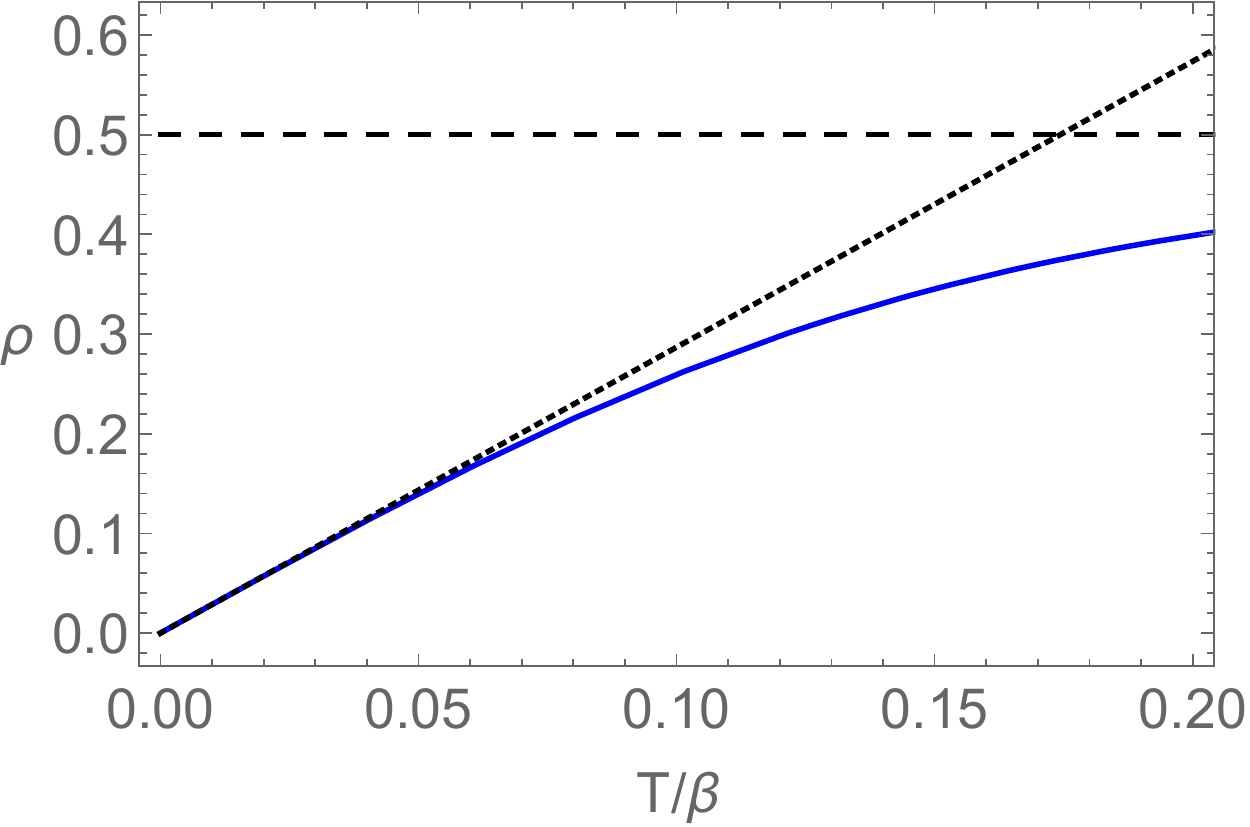} \label{}}
          \caption{ Resistivity vs temperature at fixed $\hm = 0, 0.1,1$. The horizontal dashed lines are \eqref{sigmabigTbeta} and the dotted lines are \eqref{sigmasmallTbeta}. } \label{fig7}
\end{figure}

We make a resistivity (the inverse of \eqref{conductbetafix}) plot  for $\hm=0.1, 1, 10$ in Fig. \ref{fig7}. 
The two guide lines, dotted lines and dashed lines, are the inverse of \eqref{sigmasmallTbeta} and \eqref{sigmabigTbeta} respectively. 
As $\hm$ increases (momentum relaxation becomes weaker compared to chemical potential), the resistivity curves move away from two guide lines. 
For small $\hm$, $\rho \sim 2\sqrt{2} \pi \hT$, \eqref{sigmasmallbbeta} is a good approximation for resistivity up to $\hT \sim 1/2\sqrt{2} \pi$.


\bibliographystyle{JHEP}

\providecommand{\href}[2]{#2}\begingroup\raggedright\endgroup

\end{document}